\documentclass[letterpaper,11pt]{article}

\usepackage{jheppub} 

\usepackage{subfig}
\usepackage{float}

\def\ben{\begin{equation}}
\def\een{\end{equation}}
\def\bea{\begin{eqnarray}}
\def\eea{\end{eqnarray}}

\def\dt{\delta t}

\def\nn{\nonumber}
\def\trho{{\tilde{\rho}}}

\renewcommand{\d}{\delta}

\renewcommand{\t}{\tau}		

\title{Quantum Quench in $c=1$ Matrix Model and Emergent Space-times}

\author{Sumit R. Das,}
\author{Shaun Hampton,}
\author{Sinong Liu.}

\affiliation{Department of Physics and Astronomy, University of Kentucky, Lexington, KY 40506, U.S.A.}

\emailAdd{das@pa.uky.edu}
\emailAdd{sha444@uky.edu}
\emailAdd{sinong.liu@uky.edu}

\abstract{We consider quantum quench in large-N singlet sector quantum
  mechanics of a single hermitian matrix in the double scaling
  limit. The time dependent parameter is the self-coupling of the
  matrix. We find exact classical solutions of the collective field
  theory of the eigenvalue density with abrupt and smooth quench profiles which
  asymptote to constant couplings at early and late times, and with the system initially in its ground state. With adiabatic initial conditions we find that adiabaticity is always
  broken regardless of the quench speed. In a class of
  quench profiles the saddle point solution for the collective field
  diverges at a finite time, and a further time evolution becomes ambiguous. However the underlying matrix model expressed in terms of fermions predict a smooth time evolution across this point. By studying fluctuations around the saddle point solution we interpret the emergent space-times. They generically have spacelike boundaries where the couplings of the fluctuations diverge and the semi-classical description fails. Only for very finely tuned quench profiles, the space-time is normal.}

\begin{document}


\maketitle
\flushbottom

\section{Introduction}
\label{sec:intro}

In the usual AdS/CFT correspondence, quantum quench \footnote{In this
  paper we will use "quantum quench" to describe a time dependent
  coupling, typically with a finite rate of change} in a
non-gravitational field theory is described as a time dependent
boundary condition in the gravitational dual. When the system is
initially in a nice state (e.g. the ground state) and the time dependence causes an energy injection over a
finite time period, the usual outcome is black hole formation in the
bulk \cite{bh1}-\cite{bh5a}. This manifests as thermalization in the field theory. Such quantum quenches have been accordingly used extensively to investigate thermalization in strongly coupled theories. In other situations where the quench takes the system across, or approaches a critical point this holographic connection has been used to understand universal scaling and dynamical phase transitions \cite{holo-kz}-\cite{fast}.

There are, however, other examples where holographic quenches lead to time dependent bulk backgrounds which resemble cosmologies \cite{cosmo} -\cite{cosmo5} \footnote{In these examples the time dependence turns off in the inifinite past but in an exponential fashion, rather than at some finite past time}. When the coupling of the boundary theory goes through a small value, the bulk develops a space-like or null region of high curvature, physically resembling a cosmological singularity. In this region the classical gravity approximation fails and it is interesting to ask if the dual field theory can provide an unambigious time evolution. While there are indications that {\em some} of these examples may indeed involve smooth time evolutions (especially for a null time dependence or a sufficiently slow time dependence), the nature of the final state is not clear : this possibly contains a black hole. Neverthless these examples have been useful in understanding boundary signatures of bulk singularities.

In this paper we will investigate quantum quench in the earliest
example of holography which predates the AdS/CFT correspondence :
duality of double scaled Matrix Quantum Mechanics and two dimensional
non-critical string theory \cite{cone}. This is a model of holography
where both sides of the duality are understood quantitatively : as
such this could have lessons for the general question discussed
above. The singlet sector of this model can be written exactly in
terms of a theory of non-relativistic fermions in an inverted hamonic
oscillator potential \cite{conefermion,c1fermion} and at large $N$ it is more
useful to write the
model as a collective field theory of the density of
eigenvalues \cite{jevsak}. The space on which these fermions live is the space of eigenvalues of the matrix. The fluctuations of the collective field around the large $N$ saddle point solution is a massless scalar which is related to the single dynamical degree of freedom of two dimensional string
theory \cite{dasjev}. The saddle point itself is related to a classical tachyon background of the string theory \cite{pol1}. In fact a precise relationship involves a spatial transformation whose kernel is nonlocal at the string scale \cite{polnat}. 
There is no propagating graviton, and the emergent space and gravitational effects are figured out by examining the propagation and scattering of collective field fluctuations. Recent work has made the relationship between two dimensional string theory in worldsheet formalism and the matrix model results even more precise \cite{xiyin}. The way holography works in this case is somewhat different from the usual AdS/CFT correspondence, even though the origins come from the physics of D-branes in either case \cite{vermc}. In particular this not really a "bulk-boundary correspondence". Nevertheless the description in terms of the collective field in $1+1$ dimension is a "bulk" description, while the description in terms of $N^2$ matrix degrees of freedom is the analog of the boundary description.

In two dimensional string theory, gravity and the higher string modes are not propagating degrees of freedom. Nevertheless there can be distinct {\em backgrounds} with different metrics and other fields. Indeed, the low energy description in terms of dilaton gravity has a well known black hole background and there is worldsheet theory in this background \cite{mswwitten} - this led to significant insights in the issue of black hole evaporation \cite{2dbh}. Despite many decades of effort, this black hole background is not understood in the matrix model description. It is generally believed that the singlet sector in fact does not have a black hole. One evidence for this belief is that an incoming tachyon pulse does not produce a black hole \cite{kms}. The non-formation of black holes is related to the presence of an infinite number of $W_\infty$ charges in the theory. It has been speculated that the non-singlet sector does contain a black hole \cite{kkk}- however the status of this speculation is not clear at the moment. This naturally raises the question : what happens when we subject the matrix model (restricted to the singlet sector) to a quantum quench ?

We perform quantum quench in this model by introducing a {\em time dependent self-coupling} of the matrix
which interpolates between constant values at early and late times. As
we will show, in the singlet sector and in the double scaling limit, this model maps to a system of $N$ fermions in an inverted
harmonic oscillator potential with a time dependent coefficient of the potential (We will call this coefficient
``frequency" below). Using standard methods, the collective field
description can be written down. 

We show that the action of the collective field theory with time
dependent frequency can be mapped
to one with a constant frequency plus a boundary term by a
transformation involving a single function of time $\rho (t)$ which
satisfies a generalization of nonlinear Ermakov-Pinney (EP) equation
\cite{EP}. Thus, in the classical limit, the response of the
collective theory can be obtained once we know the solution of the EP
equation with appropriate initial conditions. Actually what determines
the solution is $\rho^2(\tau)$ and this quantity can be both positive
and negative. The EP equation can be solved in terms of the independent solutions of the {\em classical} equations of motion of a single particle moving in the inverted harmonic potential with a time dependent frequency $f(t)$. 
Our method is adapted from known methods of solving the single particle Schrodinger equation in the presence of a harmonic potential with a time dependent frequency\cite{hotd}. This technique has been used to understand aspects of the quantum quench problem for $N$ non-relativistic fermions in such a potential \cite{ruggiero,dhl} as well for quench in a harmonic chain in \cite{ghosh}.

We discuss exact solutions for several quench protocols. These include abrupt quenches with piecewise constant frequencies, where the frequency increases or decreases in a single step, and a "pulse" or "dip" where the frequency rises or dips down for a finite period of time, and becomes a constant after that. For smooth quenches, our solvable examples include those which monotonically interpolate between constant values at early and late times with a time scale $\dt$, as well as cases where the frequency starts with some constant value and ends up at the same constant value after dipping down for some time.
The quench starts off with the system in a
ground state at large negative times where the energy levels on both sides of the inverted
harmonic potential are occupied upto some fermi energy.

The initial conditions are such that the collective field and its time derivative match the adiabatic solution in the far past. The initial time evolution is then adiabatic. The collective field has a cut (i.e. it vanishes inside a finite interval in eigenvalue space) which changes with time. However, in contrast to the right side up harmonic oscillator, we find that with these initial conditions adiabaticity is always broken at some finite time, regardless of the speed of the quench. This can be qualitatively understood by casting the generalized EP equation in terms of a potential problem : this analog potential is unstable which renders an adiabatic expansion invalid. As a result, the qualitative aspects of the solution are well approximated by similar profiles with abrupt changes. 

We show that generically at late times the function $\rho^2 (\tau)$  becomes infinitely large positive or negative. However, there are {\em finely tuned} pulse or dip protocols for which $\rho (\tau)^2$ approaches a constant value.

For the case where $\rho^2$ goes to negative infinity, it crosses zero at a {\em finite time} $\tau = \tau_0$ : at this time the saddle point solution for the collective field diverges, and the equations cannot predict a further time evolution unambigously. This
is therefore a situation where the ``bulk'' equations of motion fail.  However, the underlying fermion description remains well-defined even at the classical level and  predicts a smooth time
evolution across this time. For $\tau > \tau_0$ the fermions in the initial fermi sea cross
over to the other side of the potential.  Using the fermion picture
one can now define a different collective field (which is no longer the
fermion density), but nevertheless obeys the same equations of motion. At infinite
matrix model time, this new collective field vanishes everywhere. 

We then seek a space-time interpretation of the model by considering fluctuations around the saddle point solution. The fluctuation action is that of a massless scalar field in a relativistic two dimensional space-time with couplings which are space and time dependent. Since the scalar is massless, we can read off the metric from the quadratic action only upto a conformal factor. However, this can be used to derive the Penrose diagram showing the global properties of the emergent spacetime. We do this in detail for abrupt quenches. 

For a single step quench where the frequency increases, this semi-classical space-time terminates on a spacelike boundary. At this boundary the cubic couplings of the fluctuations of the collective field diverge : this is like a space-like singularity. 
Of course what this means is that the semiclassical collective theory used to obtain the space-time interpretation fails as we approach this time. The matrix model time at this point is $\tau = \infty$. To determine if the matrix model time evolution is smooth at these late times, one would need to use an exact non-perturbative treatment of the fermionic theory, perhaps along the lines of \cite{moore}. In view of the above discussion this would be the general feature of any smooth quench which leads to an increasing $\rho(\tau)^2$.

An interesting feature of this emergent relativistic space-time is that the space of eigenvalues, $x$, does not remain a space-like coordinate for all times. Constant $x$ lines are timelike before the quench, but can become null or space-like after the quench.
The signature of the emergent metric does not change.

For a single step quench where the frequency decreases, the time $\tau = \tau_0$ discussed above becomes a null ${\cal I}^\pm$. Normally this would be the boundary of space-time. However, the matrix model provides a smooth time evolution beyond this : we therefore need to append another piece of space-time. The infinite future in matrix model time $\tau = \infty$ corresponds to a space-like line in this additional piece where the cubic couplings of the fluctuations diverge. Once again all constant $x$ lines do not remain time-like.

It is only in the very fine tuned situation where $\rho(\tau)$ asymptotes to a constant in the far future that the emergent space-time is "normal".

Time dependent solutions of the matrix model with {\em constant couplings} have been studied earlier as models of matrix cosmology \cite{ks}. These solutions are generated by the underlying $W_\infty$ algebra. It turns out that for profiles of the first class where the frequency changes suddenly from one value to another, the solutions are identical to one class of such solutions \cite{dk}. We do not understand why a quench produces precisely these kinds of states.

Quantum quench in unitary and hermitian matrix models was first investigated in \cite{mandal}, and followed up in a related recent paper \cite{mandal2}. These papers deal with abrupt quenches and address questions of relaxation to a Generalized Gibbs Ensemble and dynamical phase transitions, with several novel results. Our interest is complementary : we concentrate on the question of emergent space-time.

In section 2 we describe the model and set up the notation. Section 3 deals with the method of solving the dynamics following a general quench. Section 4 deals will explicit solutions of the collective field with various quench profiles. In section 5 we discuss the nature of the solutions in the fermion picture. Section 6 discusses the nature of the emergent space-time. Section 7 contains conclusions. The appendices provide details of some of the pertinent results.

\section{The $c=1$ Matrix Model with a time dependent coupling}

The action of our model is
\ben
S = \beta_0  \int~dt~f(t)~{\rm Tr}~ [\frac{1}{2} {\dot{M}}^2 - U(M) ]
\label{1-1}
\een
where $M$ stands for a $N \times N$ hermitian matrix, and $f(t)$ is a specified function of time which goes to constant values at early and late times. $U(M)$ is a potential which has a maximum at $M=0$, e.g.
\ben
U(M) = -\frac{1}{2}M^2 + \frac{1}{4} M^4 + \cdots
\label{1-2}
\een
If we write $\beta_0 = \frac{N}{g}$, $g$ is the 't Hooft coupling. We now define a new time variable $\tau$ by
\ben
d\tau = \frac{dt}{f(t)}
\label{1-6a}
\een
so that the action becomes
\ben
S = \beta_0 \int d\tau {\rm Tr}~ [\frac{1}{2} (\partial_\tau M)^2 -  f(\tau)^2~U(M) ]
\label{1-1a}
\een
where $f(\tau)=f(t)$. 
We will consider $f(\tau)$ which remains positive for all times. Therefore $\tau$ is a monotonically increasing function of $t$.

We will consider a gauged version of the model where the $U(N)$ symmetry of (\ref{1-1}) is gauged. Since there is no dynamics of a gauge field in $0+1$ dimension this essentially means a restriction to the singlet sector. In this sector one can replace the $N^2$ degrees of freedom with the $N$ eigenvalues $\lambda_i(t)$. The jacobian of the change of variables to $\lambda_i$ is a van der Monde determinant. One can then redefine the wavefunction by absorbing a factor of the square root of this determinant - the new wavefunction is then a Slater determinant, so that we have a theory of $N$ fermions moving in the space of eigenvalues in the presence of an external potential. The second quantized hamiltonian of the fermion field $\chi(\lambda,t)$ is given by
\ben
H = \int d\lambda \left[ \frac{1}{2\beta_0}|\partial_\lambda \chi|^2 + \beta_0 f(\tau)^2 U(\lambda) |\chi|^2 + \beta_0 \mu_F |\chi|^2\right] - \beta_0 \mu_F N
\label{1-3}
\een
where we have used a Lagrange multiplier $\mu_F$ to impose the
constraint which sets the total number of fermions to $N$,
\ben
\int d\lambda |\chi|^2 = N
\label{1-3a}
\een
The double scaling limit of this model is then defined by
\ben
\beta_0 \rightarrow \infty~~~~~\mu_F \rightarrow 0~~~~~~g_s = \frac{1}{2\beta_0 \mu_F} = {\rm fixed}
\label{1-4}
\een
In this limit the model simplifies considerably. This is seen by rescaling
\ben
\lambda = (\beta_0 g_s)^{-1/2} x~~~~~~~~\chi = (\beta_0 g_s)^{1/4} \psi
\label{1-5}
\een
and retaining the $O(1)$ terms. 
The final double-scaled hamiltonian is
\ben
H = \int dx \left[ \frac{g_s}{2}|\partial_x \psi|^2-\frac{f(\tau)^2}{2g_s}x^2 |\psi|^2  + \frac{1}{2g_s} |\psi|^2 \right]
\label{1-6}
\een 
where we have ignored a constant additive term.
The``Planck constant" of the problem is $g_s$. 

If we think of this arising from e.g. a potential as in (\ref{1-2}), the quartic terms become  $O(1)$ when $x \sim \mu_F^{-1/2}$. Since $g_s$ is held fixed this corresponds to $x \sim \sqrt{N}$. Because of this, a regulated version of this model can be written down with a hard wall at $|x| \sim \sqrt{N}$.

Alternatively the singlet sector of the matrix model can be expressed in terms of a collective field $\rho(x,\tau)$ which is the density of eigenvalues, or the fermion density in eigenvalue space
\ben
\rho(x,\tau) = \partial_x \zeta (x,\tau) = {\rm Tr} \delta (M(\tau) - xI) = \psi^\dagger \psi (x,\tau)
\label{1-7}
\een
The dynamics of $\zeta (x,\tau)$ can be derived using the method of \cite{jevsak,dasjev}. For small $g_s$ one can alternatively use the classical bosonization relations of \cite{polbos}. The action is
\ben
S = \frac{1}{g_s^2} \int dx d\tau~\left[ \frac{1}{2}\frac{(\partial_\tau \zeta)^2}{\partial_x \zeta} - \frac{\pi^2}{6} (\partial_x\zeta)^3 + \frac{1}{2} [ f(\tau)^2 x^2 -1](\partial_x\zeta) \right]
\label{1-8}
\een
In the time independent situation, $f(\tau)=\omega_0$, the ground state classical solution is given by
\ben
\partial_x \zeta_0 (x) = \frac{\omega_0}{\pi}\left[ x^2 -\frac{1}{\omega_0} \right]^{1/2} ~~~~~|x| \leq 1
\label{2-4}
\een
and zero in the interval $-\frac{1}{\sqrt{\omega_0}} \leq x \leq \frac{1}{\sqrt{\omega_0}}$. Fluctuations of the collective field becomes related to the``massless tachyon" of two dimensional string theory. The coordinate $x$ which arose out of the space of eigenvalues plays the role of space.

\section{Response to a Quantum Quench}

We aim to find solutions of the equations of motion with appropriate initial conditions for a given quench profile $f(\tau)$. This is facilitated by a remarkable property of the theory. Consider a transformation of $(x,\tau) \rightarrow (y,T)$
\ben
y = \frac{x}{\rho(\tau)}~~~~~~~~~~T = \int^\tau \frac{d\tau^\prime}{\rho(\tau^\prime)^2}
\label{2-1}
\een
where $\rho(\tau)$ is some function, 
under which $\zeta$ transforms as a scalar,
\ben
\partial_\tau \zeta = \frac{1}{\rho^2} \partial_T \zeta - y \frac{\partial_\tau \rho}{\rho} \partial_y \zeta
\label{2-1a}
\een
If the function satisfies the nonlinear equation
\ben
\frac{d^2\rho}{d\tau^2} - f^2(\tau) \rho = - \frac{1}{\rho^3}
\label{2-3}
\een
the action in (\ref{1-8}) then becomes
\bea
S & = & \frac{1}{g_s^2} \int dy dT~\left[ \frac{1}{2}\frac{(\partial_T \zeta)^2}{\partial_y \zeta} - \frac{\pi^2}{6} (\partial_y\zeta)^3 + \frac{1}{2} (y^2 -1)(\partial_x\zeta) \right] + \nonumber \\
& & 
\frac{1}{g_s^2}\int dy dT \left[ \partial_y \bigg(-{1\over2}y^2 \zeta + {1\over2} f(\t)^2y^2\rho^4 \zeta  + {1\over2}y^2\rho^2(\partial_{\t}\rho)^2  \zeta\bigg) - \partial_T\bigg(y\rho\partial_{\t}\rho\zeta\bigg)\right]
\label{2-2}
\eea
This means the equations of motion map to those with a constant frequency $f=1$, so that a solution of the equations of motion of the action with a constant frequency can be lifted to a solution of the equations of motion with a time dependent frequency using a solution of (\ref{2-3}).
The equation (\ref{2-3}) is a generalization of Ermakov-Pinney equation \cite{EP}. The latter has a plus sign in front of $f^2$ and $1/\rho^3$. 

As we will see soon, the function which appears in our discussion is actually $\rho^2 (\tau)$, and this can be a both positive and negative real quantity. It is therefore useful to consider the equation for this quantity, 
\ben
\partial_{\t}^2 \rho^2 - {1\over2\rho^2}(\partial_{\t}\rho^2)^2 - 2\omega(\t)^2\rho^2= - {2\over\rho^2}
\label{2-3a}
\een
In fact we will find that it is {\em necessary} to have negative values of $\rho^2$. It is therefore useful to define the quantity
\ben
\trho (\tau) \equiv +\sqrt{|\rho(\tau)^2|}
\label{2-3bb}
\een
The rescaling involved is then really
\ben
y = \frac{x}{\trho (\tau)}
\label{2-3cc}
\een

To get an intuition about the solutions, it is useful to rewrite the generalized EP equation (\ref{2-3}) as the equation of motion of a particle in a potential,
\ben
\frac{d^2\rho}{d\tau^2} = - \frac{\partial V(\rho, \tau)}{\partial \rho}
\label{2-3b}
\een
where the potential is
\ben
V(\rho, \tau) = -\frac{1}{2}\left[f(\tau)^2 \rho^2 + \frac{1}{\rho^2} \right]
\label{2-3c}
\een
When $f(\tau) = \omega_0$ for all $\tau$ the ground state solution is given by (\ref{2-4}). This means that in this case we need to choose a constant solution of the generalized EP equation (\ref{2-3}), $\rho^2 = \frac{1}{\omega_0}$. 

The quench profile we are interested in asymptotes to a constant value $\omega_0$ at $t \rightarrow -\infty$. Therefore if we start out the system in its ground state we need to solve the equation (\ref{2-3}) such that it asymptotes to a constant value at early times.
We will in fact use profiles which are either piecewise constant or become constant exponentially at early and late times. Thus the time evolution near $\tau= -\eta$ for large enough positive $\eta$ should be adiabatic. This means we need to find solutions of (\ref{2-3}) which match on to the adiabatic solution
\ben
\rho_{ad}(\tau)^2 = \frac{1}{f(\tau)}
\label{2-5}
\een
at some very early time. These conditions are, for a large negative $T$
\ben
\rho^2(T) = \frac{1}{f(T)}~~~~~\partial_\tau \rho^2 (\tau) = - \frac{\partial_\tau f(\tau)}{f(\tau)^2}|_{\t=T}
\label{2-6}
\een

Given such a solution, the  time dependent classical solution for the original action can be easily written down using (\ref{2-1})
\bea
\partial_x \zeta_0 (x,\tau) & = & \frac{1}{\pi \rho(\tau)^2}\left[ x^2 -\rho(\tau)^2 \right]^{1/2} \\
\partial_\tau \zeta_0 (x,\tau) & = & - \frac{\partial_\tau \rho (\tau)^2}{2 \rho (\tau)^2} x \partial_x \zeta (x,\tau) 
\label{2-6a}
\eea
It can be easily checked that these satisfy the consistency condition
\ben
\partial_x\partial_\tau \zeta_0 (x,\tau) = \partial_\tau\partial_x \zeta_0 (x,\tau)
\label{2-6b}
\een

There is a well known way to find solutions of the EP equation which we adapt to the generalized equation \cite{hotd}. The most general solution of  (\ref{2-3}) is given by
\ben
\rho (\tau)^2 = A u(\tau)^2 + 2B u(\tau)v(\tau) + C v(\tau)^2
\label{2-7}
\een
where $A,B,C$ are constants and $u(\tau), v(\tau)$ are two linearly independent solutions of the {\em classical} equation of motion of a single particle moving in an inverted harmonic potential with the same time dependent frequency $f (\tau)$
\ben
\partial_\tau^2 X - f(\tau)^2 X = 0
\label{2-8}
\een
Furthermore $A,B,C$ must satisfy
\ben
AC - B^2 = - \frac{1}{Wr(u,v)^2}
\label{2-9}
\een
where $Wr(u,v) = u \partial_\tau  v - v \partial_\tau u$ is the wronskian of the two solutions. By the equations of motion this is a constant in time and can be therefore evaluated at any time. 

Given a classical solution of the action (\ref{1-8}), $\zeta_0(x,t)$, the next step is to obtain the action for small fluctuations by expanding
\ben
\zeta (x,t) = \zeta_0 (x,t) +\frac{ g_s}{\sqrt{\pi}} \eta(x,t)
\label{2-10}
\een
The action for the fluctuations will clearly be nonpolynomial in $\eta$. The quadratic part of the fluctuation action is
\ben
S^{(2)} = \frac{1}{2\pi} \int dx d\tau \left[ \frac{(\partial_\tau \eta)^2}{\partial_x\zeta_0}
-2 \frac{(\partial_\tau \zeta_0)}{(\partial_x \zeta_0)^2} (\partial_\tau \eta)(\partial_x\eta) + \left( \frac{(\partial_\tau \zeta_0)^2}{(\partial_x \zeta_0)^3} - \pi^2 \partial_x\zeta_0 \right) (\partial_x \eta)^2 \right]
\label{2-11}
\een
while the cubic interaction part is 
\ben
S^{(3)} = -\frac{1}{\pi^{3/2}}\int dx d\tau \left[ \frac{1}{2} \frac{1}{(\partial_x \zeta_0)^2} (\partial_\tau \eta)^2 (\partial_x \eta)- \frac{\partial_\tau \zeta_0}{(\partial_x \zeta_0)^3}(\partial_\tau \eta)(\partial_x \eta)^2 + \left( \frac{(\partial_\tau \zeta_0)^2}{(\partial_x \zeta_0)^4} + \frac{\pi^2}{6} \right) (\partial_x \eta)^3 \right]
\label{2-12}
\een
The quadratic action shows that the fluctuation field is a {\em relativistic} massless field which is propagating on a $1+1$ dimensional space-time with a metric which is conformal to 
\ben
ds^2  =  -d\tau^2 + \frac{(dx + \frac{\partial_{\tau} \zeta_0}{\partial_x \zeta_0}d \tau)^2}{(\pi \partial_x \zeta_0)^2} 
\label{2-13}
\een
Since we are dealing with a massless field in $1+1$ dimensions the quadratic action is insensitive to a conformal transformation of the metric. Note that the signature of the metric (\ref{2-13}) remains negative at all times since 
${\rm det}(g) = -\frac{1}{\pi^2(\partial_x \zeta)^2}$.

\section{Solutions for some quench profiles} 
In this section we find solutions of the generalized EP equation for physically interesting quench profiles.

\subsection{ Abrupt Quenches} 
Consider first abrupt quenches. The first case we consider is a single step quench, where the function $f(\tau)$ appearing in e.g. (\ref{1-8}) is given by
\begin{equation}
f(\tau) = \left\{
  \begin{array}{lr}
    \omega_0 &:  \tau < 0 \\
   \omega_1 &:  \tau > 0
  \end{array}
\right.
\label{abrupt}
\end{equation}

\begin{figure}[!h]
\centering
\includegraphics[width=2.5in]{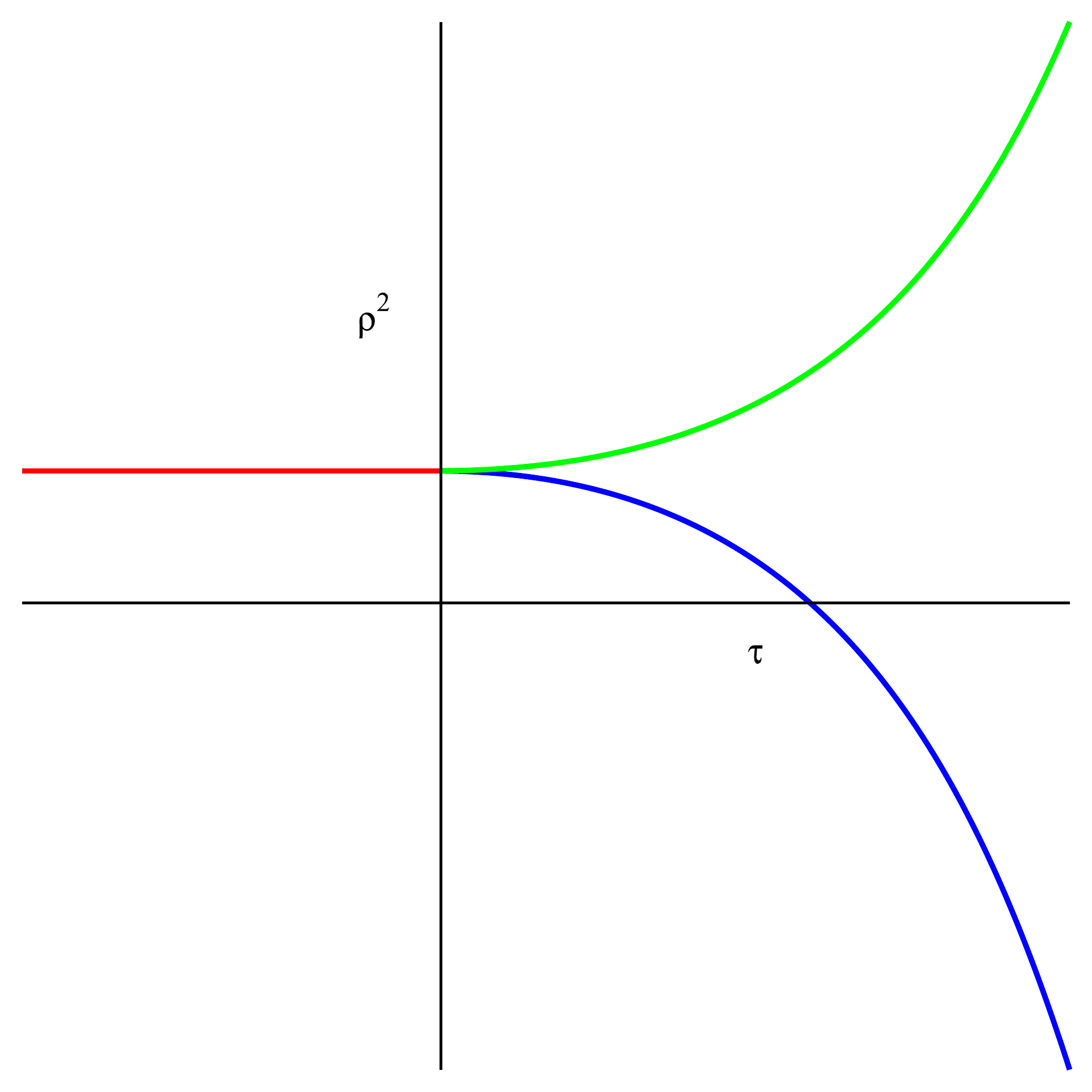}
\caption{$\rho^2(\tau)$ for abrupt quench. We have chosen $\omega_0=1$. The green line has $\omega_1=\sqrt{2}$ while the blue line has $\omega_1=1/\sqrt{2}$}
\centering
\label{fig-one}
\end{figure}

In this case the two linearly independent solutions to the classical equations of motion (\ref{2-8}) may be chosen to be
\bea
u(\tau) & = &\theta(-\tau) e^{\omega_0 \tau} + \theta(\tau) \frac{1}{2}\left[
(1+ \frac{\omega_0}{\omega_1})e^{\omega_1\tau} +
(1- \frac{\omega_0}{\omega_1})e^{-\omega_1\tau} \right] \nonumber\\
v(\tau) & = & \theta(-\tau) e^{-\omega_0 \tau} + \theta(\tau) \frac{1}{2}\left[
(1- \frac{\omega_0}{\omega_1})e^{\omega_1\tau} +
(1+ \frac{\omega_0}{\omega_1})e^{-\omega_1\tau} \right]
\label{3-1a}
\eea
Using (\ref{2-7}) and (\ref{2-9}) we need to find a solution for $\rho^2$ which is $\frac{1}{\omega_0}$ for $\tau < 0$. This clearly requires a choice $A=C=0$ in (\ref{2-7}). Then $(\ref{2-9})$ requires $B = \frac{1}{\omega_0}$. This leads to the solution for $\rho^2(\tau)$
\ben
\rho^2(\tau) =\frac{1}{\omega_0} \left[ \theta(-\tau) + \theta(\tau) \cosh^2\omega_1\tau \left(1 - \frac{\omega_0^2}{\omega_1^2} \tanh^2 \omega_1\tau \right) \right]
\label{3-2}
\een
The function $\rho^2(\tau)$ is shown in Figure \ref{fig-one}. This monotonically increases for $\omega_0 < \omega_1$, while for $\omega_0 > \omega_1$ this montonically decreases, crossing a zero at a finite time $\tau_0$ given by
\ben
\tanh \omega_1 \tau_0 = \frac{\omega_1}{\omega_0}
\label{3-3}
\een
At this point the saddle point solution $\partial_x\zeta_0$ (\ref{2-6a}) diverges. Note that $\partial_\tau \rho^2$ is finite here - this implies that the ratio $(\partial_\tau \zeta_0)/(\partial_x \zeta_0)$ diverges as well.

\begin{figure}[!h]
\centering
\includegraphics[width=2.5in]{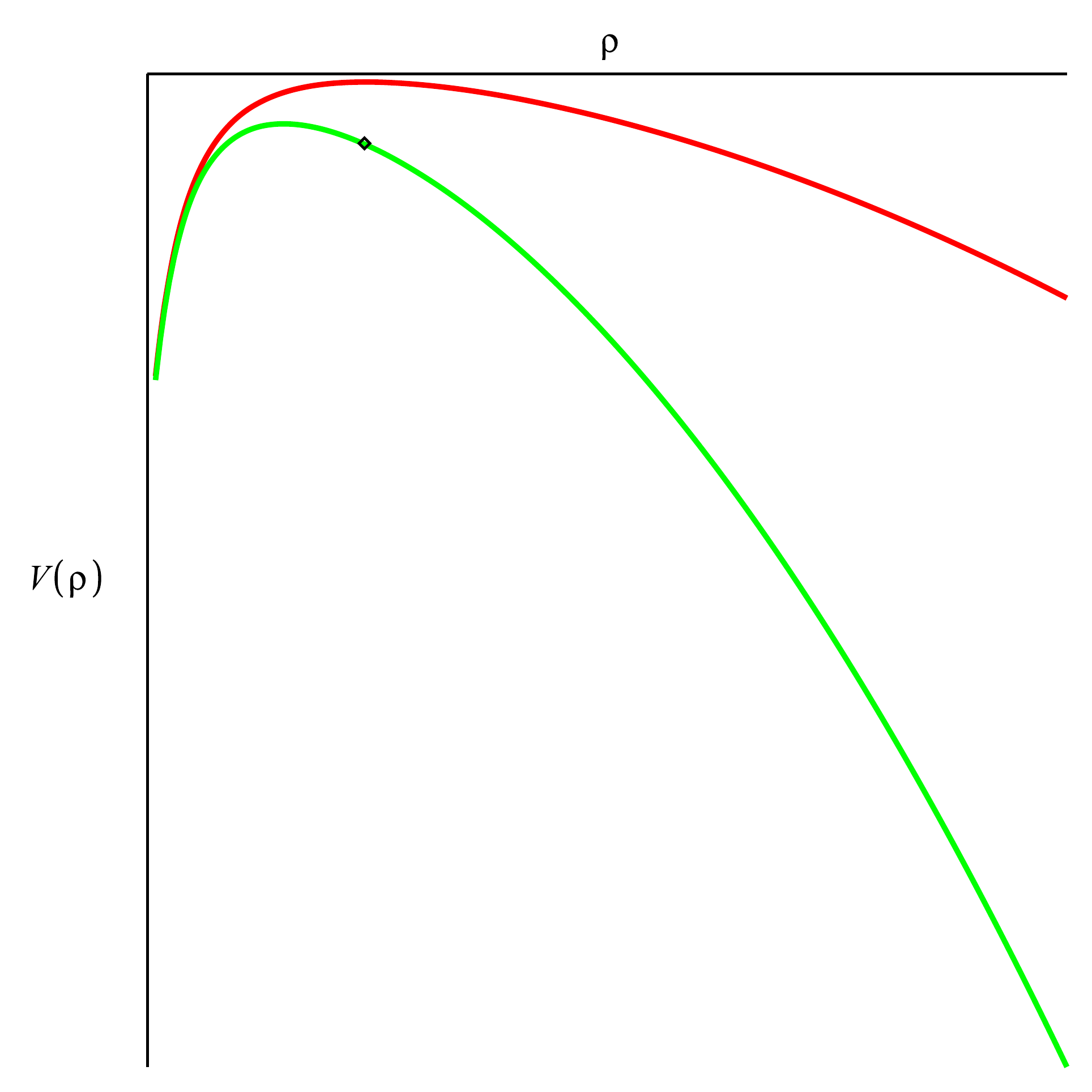}
\caption{The potential in the Ermakov-Pinney analog potential problem for abrupt quench for $\omega_0 < \omega_1$ . The red curve is the potential for $\tau < 0$, while the green curve is the potential for $\tau > 0$. The black dot is the position of the analog particle at $\tau=0$.}
\label{fig-two}
\end{figure}

The behavior of $\rho(\tau)$ can be understood from the analog potential $V(\rho)$ (\ref{2-3c}) which appears in the generalized EP equation. In Figure \ref{fig-two} the red curve is the potential for $\tau < 0$, the green curve is the potential for $\tau > 0$ in the case $\omega_0 < \omega_1$.  For $\tau < 0$ our initial conditions mean that the analog particle starts off at the maximum of the red potential, $\rho = \frac{1}{\sqrt{\omega_0}}$. Thus the initial position is to the right of the maximum of the new potential and therefore the particle rolls down to large $\rho$, reaching $\rho = \infty$ at $\tau = \infty$.

\begin{figure}[!h]
\centering
\includegraphics[width=2.5in]{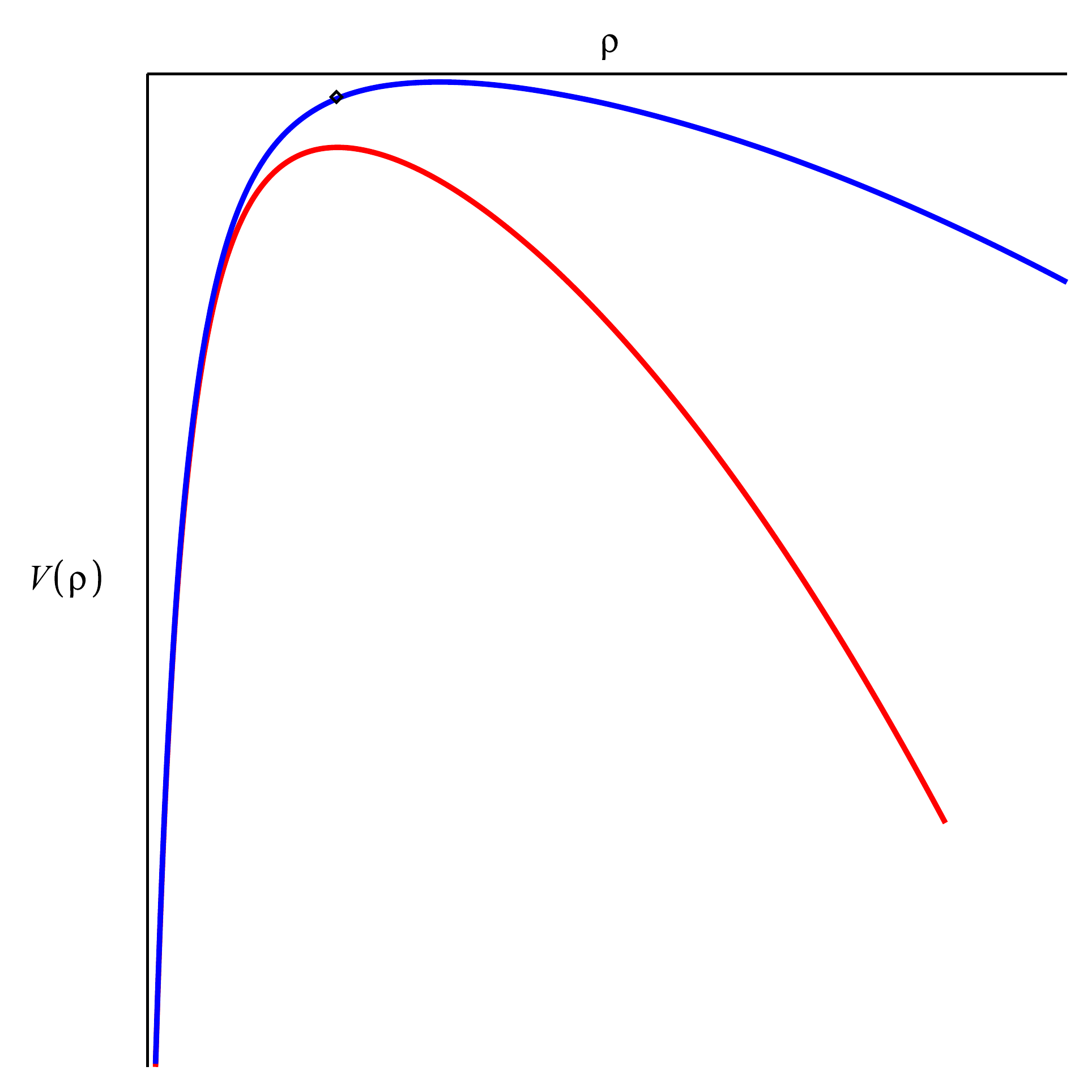}
\caption{The potential in the Ermakov-Pinney analog potential problem for abrupt quench for $\omega_0 > \omega_1$. The red curve is the potential for $\tau < 0$, while the green curve is the potential for $\tau > 0$. The black dot is the position of the analog particle at $\tau=0$.}
\label{fig-three}
\end{figure}

In Figure \ref{fig-three} the red curve is the potential for $\tau < 0$, the blue curve is the potential for $\tau > 0$ in the case $\omega_0 > \omega_1$.  Now the initial position is to the left of the maximum of the new potential and therefore the particle rolls down towards $\rho = 0$. It may be easily checked that it reaches $\rho = 0$ at a finite time $\tau = \tau_0$. This analog problem does not tell us what to do after this time. However, as explained above, the relevant quantity is not $\rho(\tau)$ itself, but $\rho(\tau)^2$. The solution (\ref{3-2}) continues to hold for $\tau > \tau_0$ and satisfies all the continuity requirements for the equation (\ref{2-3a}).

For quench profiles which are not monotonic, the late time behavior can be more interesting. Consider a series of abrupt quenches given by
\[f(\tau) = \left\{
  \begin{array}{lr}
    \omega_0: &  -\infty < \tau < -T/2 \\
   \omega_2: &  -T/2 < \tau  < T/2 \\
\omega_1: &  T/2 < \tau  <\infty
  \end{array}
\right.
\]
Here $\omega_0,\omega_1,\omega_2$ are positive and nonzero.
When $\omega_0 > \omega_2$ we will call this a "dip", while the $\omega_0 < \omega_2$ will be called a "pulse".

The details of the solution for $\rho^2(\tau)$ are given in appendix \ref{abrupt ccp}. The solution to $\rho^2(\t)$ is given by
\bea
\rho^2&=&\begin{cases}
 {1\over\omega_0},\qquad \t \leq -{T\over2}\\
\\
{1\over\omega_0}\big(AA'e^{2\omega_2\t}+BB'e^{-2\omega_2\t} + AB'+A'B\big),\qquad  -{T\over2}\leq\t \leq {T\over2}
\\
\\
{1\over\omega_0}\big(CC'e^{2\omega_1\t}+DD'e^{-2\omega_1\t} + CD'+C'D\big),\qquad \t \geq  {T\over2}
\end{cases}
\label{rhos 2}
\eea
where $A,A',B,B',C,C',D,D'$ are integration constants which are functions of $\omega_0, \omega_1, \omega_2$ which are given explcitly in Appendix \ref{abrupt ccp}.

\begin{figure}[!h]
\centering
\includegraphics[width=3.0in]{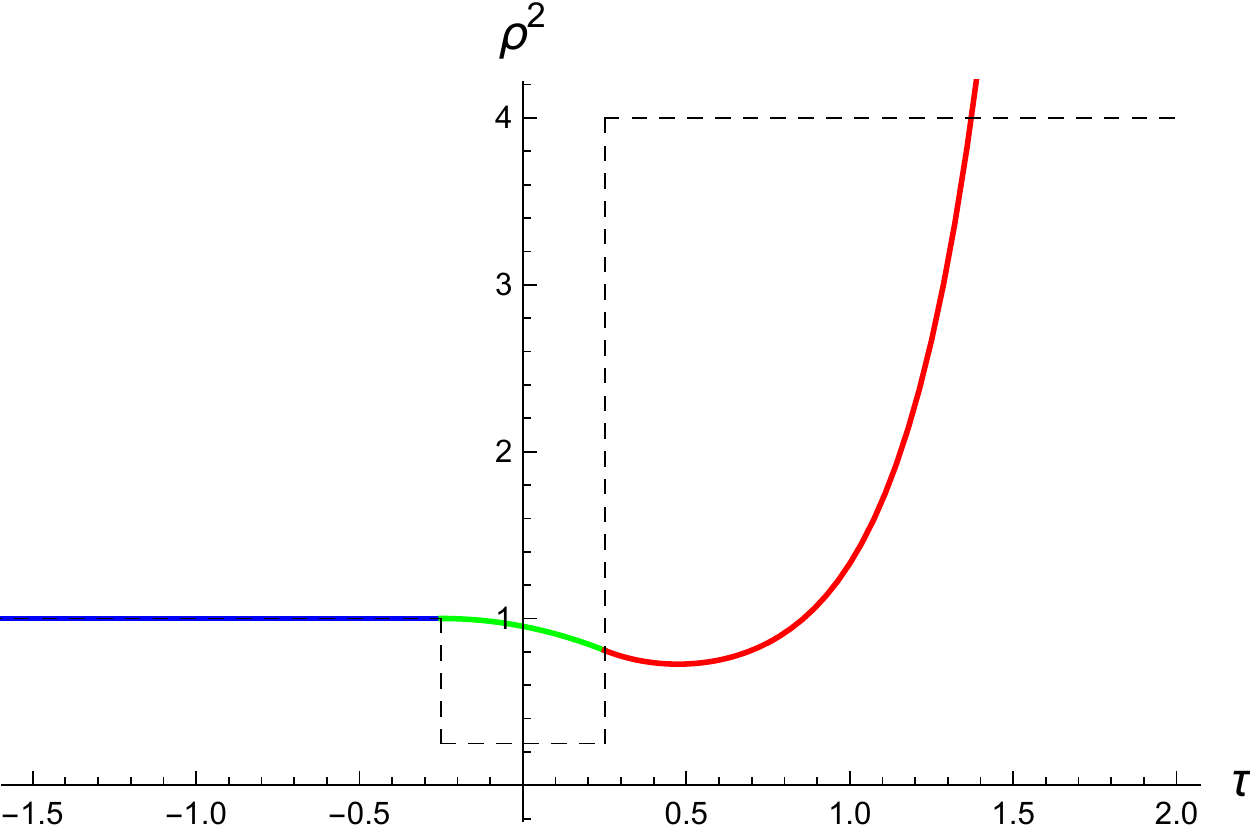}
\caption{A non-monotonic $\rho^2 (\tau)$ for an abrupt dip profile with $\omega_0=1,\omega_2=1/2$ and $\omega_1=2$ with $T=1/2$. The dashed line is the profile of $f(\tau)^2$. }
\label{nonmonotonic}
\end{figure}

In this case, $\rho(\tau)^2$ can be non-monotonic, and 
generically diverges as $\tau \rightarrow \infty$, approaching either $+\infty$ or $-\infty$.
An example of a non-monotonic solution is given in Figure \ref{nonmonotonic}.

\begin{figure}[!h]
\centering
\includegraphics[width=3.0in]{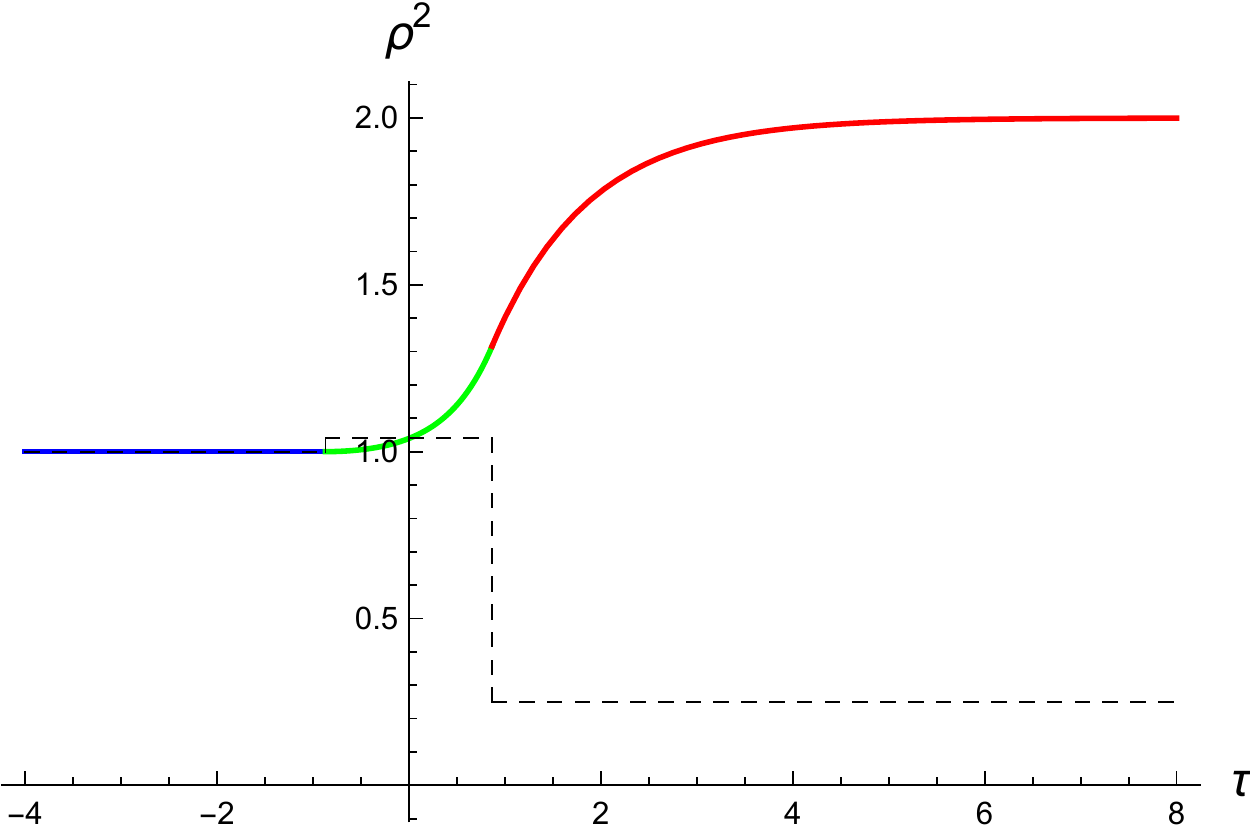}
\caption{$\rho (\tau)^2$ for a fine-tuned pulse quench with $\omega_0 = 1, \omega_2 = 51/50, \omega_1=1/2$. The blue part of the curve corresponds to $\tau < -T/2$, the green part for $-T/2 < \tau < T/2$ and the red part is $\tau > T/2$. The quench profile is shown by dashed lines.}
\label{pulserho}
\end{figure}

\begin{figure}[!h]
\centering
\includegraphics[width=3.0in]{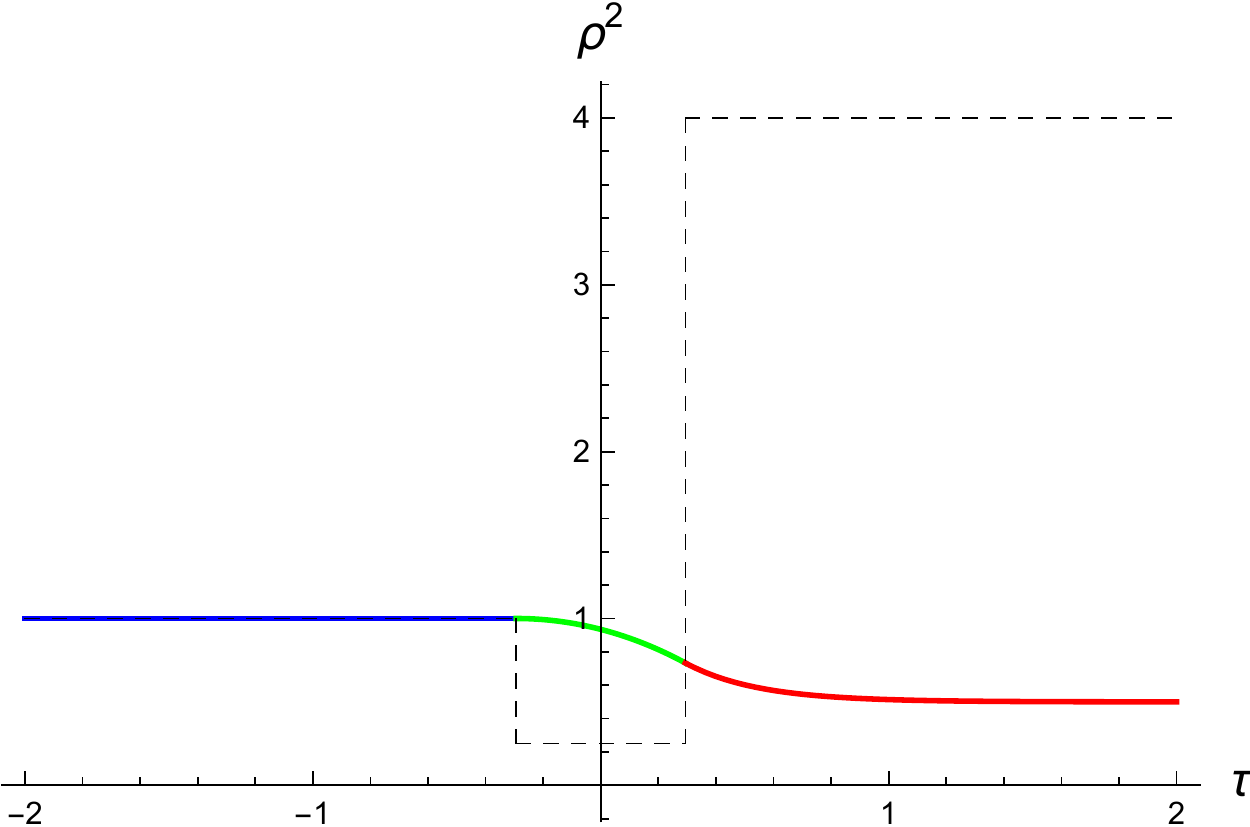}
\caption{$\rho (\tau)^2$ for a fine-tuned dip quench with $\omega_0 = 1, \omega_2 = 1/2, \omega_1=2$. The blue part of the curve corresponds to $\tau < -T/2$, the green part for $-T/2 < \tau < T/2$ and the red part is $\tau > T/2$. The quench profile is shown by dashed lines.}
\label{diprho}
\end{figure}

However there are finely tuned profiles for which $\rho^2 (\tau)$ approaches a constant at late times. The generalized EP equation of course determines this constant to be $\frac{1}{\omega_1}$.
To explore the late time behavior we need to look at the solution in the region $T/2 \leq \tau <\infty$. This is given by the last equation of (\ref{rhos 2}). For the discussion below we need the explicit forms of  $C,C'$ given below
\bea
C & = & \frac{e^{-\frac{T}{2} (2\omega_2+\omega_1+\omega_0)}}{4\omega_1\omega_2} \left[ (\omega_2 - \omega_0)(\omega_1-\omega_2) + e^{2T\omega_2} (\omega_2+\omega_1)(\omega_2+\omega_0) \right] \\
C' & = & \frac{e^{-\frac{T}{2} (2\omega_2+\omega_1-\omega_0)}}{4\omega_1\omega_2} \left[ (\omega_2 +\omega_0)(\omega_1-\omega_2) + e^{2T\omega_2} (\omega_2+\omega_1)(\omega_2-\omega_0) \right]
\eea
If $\rho(\tau)^2$ approaches a constant (which must be $\frac{1}{\omega_1}$ ) at late times either $C$ or $C'$ must vanish. It is straightforward to see that $C$ cannot vanish since $e^{2\omega_2T} > 1$, while one can find a nonzero finite value of $T$ when $C'=0$ both for a pulse $\omega_2 > \omega_0 > \omega_1$, as well for a dip $\omega_1 > \omega_0 > \omega_2$. The corresponding $\rho(\tau)^2$ are shown in Figures \ref{pulserho} and \ref{diprho}.

\subsection{Smooth Quenches}

\begin{figure}[!h]
\centering
\includegraphics[width=4.0in]{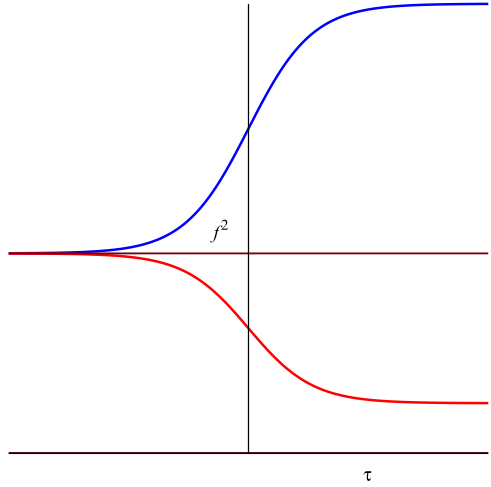}
\vspace{-2.1in}
\caption{The function $f(\tau)$. The red curve has $\omega_0 > \omega_1$ and the blue curve has $\omega_0 < \omega_1$ }
\label{quenchprofile}
\end{figure}

Consider now a smooth quench profile given by
\ben
f(\tau)^2 = \frac{\omega_1^2 + \omega_0^2 e^{-\frac{\tau}{\dt}}}{1+ e^{-\frac{\tau}{\dt}}}
\label{5-1}
\een
The quench profiles are shown in Figure \ref{quenchprofile}.

In this case the equation (\ref{2-8}) admits analytic solutions in
terms of hypergeometric functions. When $\omega_0\dt$ is not a half
integer, a choice of the two independent solutions is 
\bea
u(\tau) & = & e^{\omega_0\tau}{} _2 F_1
[\dt(\omega_0-\omega_1),\dt(\omega_0+\omega_1),1+2\dt\omega_0,-e^{\frac{\tau}{\dt}}]
\nonumber \\
v(\tau) & = & e^{-\omega_0\tau} {}_2 F_1
[-\dt(\omega_0+\omega_1),\dt(-\omega_0+\omega_1),1-2\dt\omega_0,-e^{\frac{\tau}{\dt}}]
\label{5-2}
\eea
Since the frequency approaches a constant exponentially at early times,
the intial time evolution is adiabatic. 
We need to find linear combinations of $u(\tau)$ and $v(\tau)$ in
(\ref{5-2}) such that the function $\rho
(\tau)$ constructed out of these solutions satisfy adiabatic initial
conditions at an early time. The adiabatic solutions are given by
\bea
v_{ad}(\tau) & = & \frac{C_v}{\sqrt{f(\tau)}} \exp \left[ - \int^\tau f(\tau')
  d\tau' \right] \nonumber \\
u_{ad}(\tau) & = & \frac{C_u}{\sqrt{f(\tau)}} \exp \left[ \int^\tau f(\tau')
  d\tau' \right] 
\label{5-3}
\eea
where the constants of integration $C_u,C_v$ are chosen such that as
$\tau \rightarrow -\infty$ these solutions behave as
\bea
v_{ad}(\tau) &\sim& \frac{1}{\sqrt{\omega_0}} \exp \left[ -\omega_0\tau
  \right] \nonumber \\
u_{ad}(\tau) &\sim& \frac{1}{\sqrt{\omega_0}} \exp \left[ \omega_0\tau
  \right]
\label{5-4}
\eea
Adiabatic initial conditions mean that at time $t = T$ where
$T$ is very large and negative,
\ben
\left[
\begin {array}{c}
u(T), \partial_\tau u(T)\\
\noalign{\medskip}
v(T), \partial_\tau v(T)
\end {array}
\right] \rightarrow
\left[
\begin {array}{c}
u_{ad}(T), \partial_\tau u_{ad}(T)\\
\noalign{\medskip}
v_{ad}(T), \partial_\tau v_{ad}(T)
\end {array}
\right]
\een
The specific linear combinations which satisfy this are given in the
Appendix \ref{smooth ecp}.

The function $\rho(\tau)$ is then constructed by choosing $A=C=0$ and
$B =\frac{1}{2}$ in (\ref{2-7}),
\ben
\rho(\tau)^2 = u(\tau) v(\tau)
\label{5-6}
\een
At early times, this leads to  
the correct adiabatic initial conditions (\ref{2-6}) .

An important aspect of the subsequent time evolution is that {\em regardless of the value of} $\dt$, adiabaticity is always broken. This fact can be again understood from the feature of the potential of the analog problem. 

Again, we regard Ermakov-Pinney equation as the equation of motion of an analog particle on potential (\ref{2-3c}). At a very early time, the analog particle is the maximum of the initial potential, $\rho =\frac{1}{\sqrt{\omega_0}}$. Then, at an early stage of a smooth quench, a slight variation of $f(\tau)$ causes a perturbation on the analog particle from this (unstable) equilibrium point, in a way similar to what is shown in figures \ref{fig-two} and \ref{fig-three}. Although the equilibrium point is unstable, the particle can still move back to the new equilibrium point if the initial velocity $\partial_{\tau}{\rho}$ is finely tuned, otherwise it will either pass the new equilibrium point (when $\partial_{\tau}{\rho}$ is too large), or be bounced back (when $\partial_{\tau}{\rho}$ is too small). In both cases the analog particle moves away from equilibrium point. 

Another way to see the failure of adiabaticity is to go back to the independent solutions in (\ref{5-2}). The solutions $u(\tau),v(\tau)$ describe the motion of classical particles in inverted harmonic oscillator potential $-\frac{1}{2}f(\tau)^2 x(\tau)^2$. When particles move in such a potential with a cutoff $x_b$ as boundaries, particles with negative energy $E=-\nu$ move between the boundary and the potential. Then we can figure out the adiabatic invariant of the system at $\tau \to -\infty$
\begin{equation}
I = \frac{1}{2\pi} \oint p {\text d}x = \frac{1}{2\pi} \oint \sqrt{-2\nu+\omega_0^2 x^2}{\text d} x
\end{equation}
and therefore the period of the particle is 
\begin{equation}
T =2\pi \frac{\partial I}{\partial E}=\frac{1}{\omega_0}\left[-\log 2\nu+ 2\log \left( \omega_0 x_b + \sqrt{\omega_0^2 x_b^2-2\nu} \right)\right]
\end{equation}
The adiabatic approximation holds when 
\begin{equation}
T \frac{{\text d} f (\tau)}{{\text d} \tau} \ll f (\tau)
\label{adiacond}
\end{equation} 
The solutions $u(\tau)$ and $v(\tau)$ however represent trajectories which have 
zero energy at infinite past, and these have infinitely large $T$, which violate the condition (\ref{adiacond}). Thus, the adiabatic approximation fails for $u(\tau), v(\tau)$ and therefore for $\rho(\tau)$.

There is of course a solution which is fine-tuned by specifying initial and final conditions for $u(\tau),v(\tau)$ which lead to $\rho \to \frac{1}{\sqrt{\omega_0}}$ in the infinite past and $\rho \to \frac{1}{\sqrt{\omega_1}}$ in the infinite future  
\footnote{We found the finely-tuned solutions 
\begin{equation}
\begin{split}
v_{finely-tuned} 
 =&  \left( \frac{\Gamma(1+2 \omega_1 \delta t)\Gamma(2\omega_0 \delta t)}{\delta t(\omega_0+\omega_1)\Gamma(\delta t(\omega_0+\omega_1))^2}\right)^{-1} \frac{1}{\sqrt{\omega_0}} e^{-\omega_1 \tau} {_2F_1}(-\delta t(\omega_0-\omega_1),\delta t(\omega_0+\omega_1);1+2\delta t \omega_1;-e^{-\tau/\delta t}), \\
u_{finely-tuned} =& \frac{1}{\sqrt{\omega_0}} e^{\omega_0 \tau} {_2F_1}([\omega_0-\omega_1]\delta t,[\omega_0+\omega_1]\delta t,1+2\omega_0\delta t , -e^{\tau/\delta t})
\end{split}
\end{equation}}.
The initial value of $\partial_\tau \rho$ at some $\tau = -\eta$ will be different from (\ref{2-6}). 
Only when $\delta t \gg \eta \to \infty$, $\partial_{\tau}^2 \rho \to 0$ will the adiabatic approximation become the exact finely-tuned solution given that ${\text d}\omega /{\text d}\tau \propto \delta t^{-1}$. Now, because the generalized Ermakov-Pinney equation describes a Lyapunov unstable system, according to the definition of Lyapunov stability, there exists a value $\epsilon$, for all exact solutions with initial adiabatic condition at $\tau=-\eta$ that satisfy $|\rho_{finely-tuned}(-\eta)-\rho_{ad}(-\eta)|<\delta$, where $\delta$ is a function of $\epsilon$ and time $\tau$, we can find there exists some $\tau>-\eta$ where $|\rho_{finely-tuned}(\tau)-\rho_{ad}(\tau)|>\epsilon$; i.e. at some late time, finely-tuned solution and exact solution with adiabatic initial condition are no longer close no matter how large $\eta$ is. This explains why we cannot find an exact adiabatic solution or even an exact finely-tuned solution with adiabatic initial conditions at $\tau =-\eta$ (\ref{2-6}).

\begin{figure}[!h]
\centering
\includegraphics[width=3.0in]{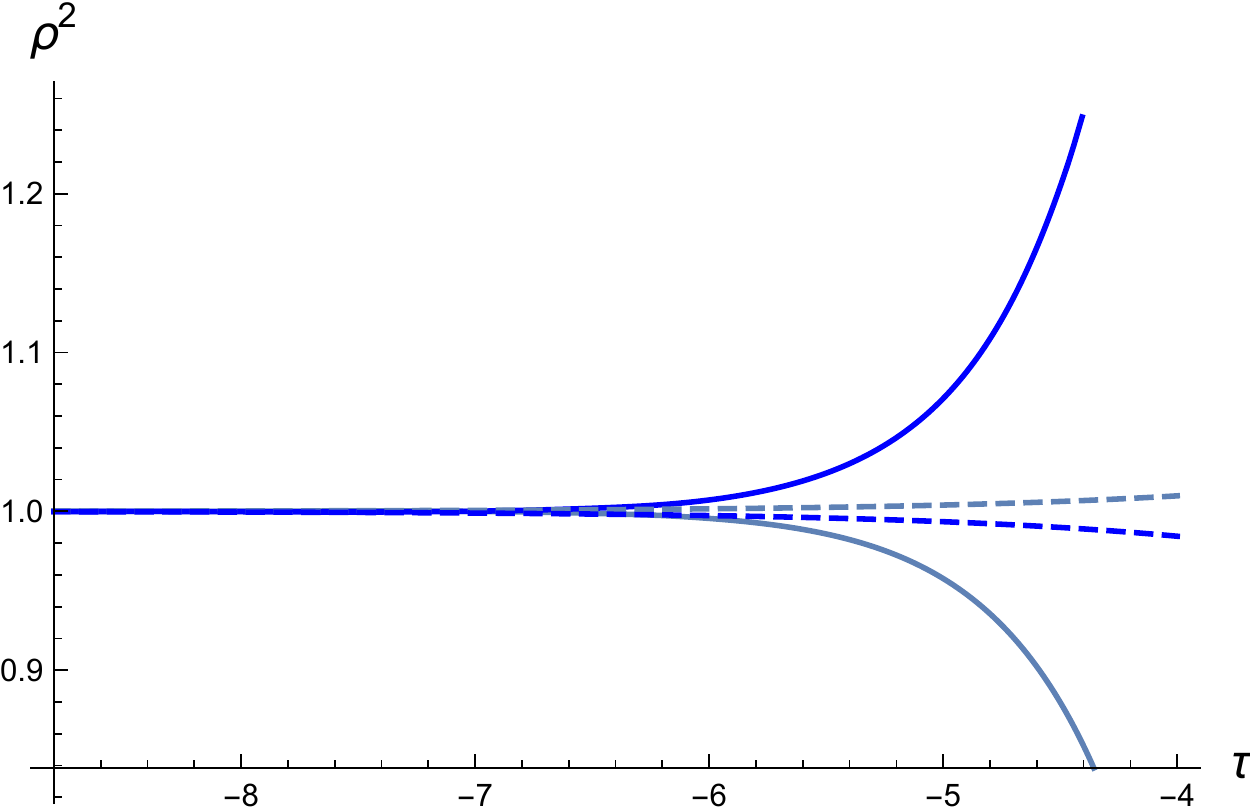}
\caption{The solution for $\rho^2(\tau)$ for the smooth profile of the form (\ref{5-1}). These have $\dt = 1.1$ and the adiabatic condition is imposed at $T = - 8.8$. The grey lines are for $\omega_0=1, \omega_1 =0.5$ while the blue lines are for $\omega_0=1, \omega_1=1.5$. The dashed lines are the adiabatic solutions, while the solid lines are the exact solutions. }
\label{ecpsmooth}
\end{figure}

Figure \ref{ecpsmooth} shows the exact and adiabatic solutions to the EP equation for the profile (\ref{5-1}), both for  $\omega_1 < \omega_0$ and $\omega_1 > \omega_0$ together with the adiabatic solution (the dashed lines). The explicit solution for this case is given in Appendix \ref{smooth ecp}.
Clearly the nature of the solutions are quite similar to those which result from abrupt quenches. In the following we will use the abrupt quench solutions to map out the emergent space-times

Another example of a solvable quench profile is give by a smooth dip or pulse,
\bea
f(\t)^2  &=& \omega_1^2+ (\omega_0^2 - \omega_1^2)\tanh^2 {\t\over\d t}
\label{smooth CCP profile}
\eea
The results for this profile are presented in appendix \ref{smooth ccp}. 

\subsection{Collective Field Saddles}

At $t \rightarrow -\infty$ the collective field in (\ref{2-6a}) vanishes in the interval $-\frac{1}{\sqrt{\omega_0}} < x < \frac{1}{\sqrt{\omega_0}}$ and monotonically increases with $|x|$. Given a solution $\rho(\tau)^2$ we can now substitute this in (\ref{2-6a}) to obtain a classical solution of the collective field. We saw that generically there can be three kinds of late time behavior for $\rho (\tau)^2$.

First, $\rho (\tau)^2$ can approach $+ \infty$ at late times. In this case, the interval of $x$ in which the collective field vanishes increases monotonically to arbitrarily large values, while the value of $\partial_x\zeta_0$ keeps decreasing till it vanishes for any finite $|x|$. This behavior is strictly for the inverted oscillator potential. For any finite but large $\beta$ (or finite $N$) this potential needs to be regulated by some kind of wall at $x \sim \sqrt{\beta g_s}$. The modification which comes from this will be clear when we discuss the solutions in the fermion language.

Secondly, $\rho (\tau)^2$ can approach $-\infty$, crossing a zero value at a {\em finite time} $\tau = \tau_0$. At this time the collective field diverges. The collective field equations cannot be used to unambigously predict a future time evolution. As we will see, one needs to go back to the fundamental fermion description to figure out if there is anything singular going on here. In fact we will see that the fermion description provides a smooth time evolution and also provide us with a continuation of the collective field.
Again, the divergence of the collective field is a feature of the strict double scaled limit. For finite large $\beta$ this becomes at most of order $\sqrt{\beta}$ - see below. In any case this region is beyond the regime where we expect the classical collective description to be good.

Finally, there can be finely tuned profiles where $\rho (\tau)^2$ approaches a constant value. In this case, the collective field obtained from (\ref{2-6a}) is finite and well defined at all times and nothing special happens. 

In the matrix model - 2d string theory duality the collective field description is the bulk description - what we are finding is that this bulk description becomes problematic except in a very finely tuned situation.

Remarkably the solutions for the single step abrupt quenches for $\tau > 0$ are exactly those which appeared earlier as time dependent solutions of the matrix model with a {\em time-independent} potential \cite{dk}. The quench solutions for other quench profiles do not correspond to such solutions in a time independent potential.

\section{The fermionic description}

The fundamental description of the theory is given in terms of fermions. 
In the semiclassical limit of small $g_s$ the fermionic theory can be understood in terms of the dynamics of the filled fermi sea in the single particle phase space. In this regime the density in phase space $u(x,p,\tau)$ is either $1$ or zero, and satisfies the Euler equation
\ben
\left[ \partial_\tau + p\partial_x + f(\tau)^2 x\partial_p \right] u(x,p,\tau) = 0
\label{4-4}
\een
There is a canonical transformation in the phase space
\ben
 y = \frac{x}{\rho(\tau)}~~~~T = \int^\tau \frac{d\tau'}{\rho (\tau')^2}~~~~P = \rho(\tau) p -(\partial_\tau \rho) x
\label{4-5}
\een
which transforms this equation to the one for a constant unit frequency
\ben
\left[ \partial_T + P \partial_y +  y \partial_P \right] u(y,P,T) = 0
\label{4-6}
\een
This may be used to write down the expression for the boundary of the filled fermi sea which corresponds to our solution for the collective field
\ben
x^2 - \big(\rho(\tau)^2 p - \frac{x}{2} \partial_\tau \rho^2 \big)^2 = \rho(\tau)^2
\label{4-7}
\een
We have expressed this entirely in terms of $\rho^2$ since this is the quantity which appears in the solutions. There are two fermi seas corresponding to the two sides of the potential. The upper and lower edges of the fermi sea for a given value of $x$ are then given by
\ben
p_\pm (x,\tau) = \frac{1}{\rho(\tau)^2} \left[ \frac{x}{2} \partial_\tau \rho^2 \pm \sqrt{x^2 -\rho (\tau)^2} \right]
\label{4-8}
\een
In the following we will denote the points on the left branch by $p_\pm^<$ and the right branch by $p_\pm^>$.

The quantities $p_\pm (x,\tau)$ are related to the collective fields by the classical bosonization relations \cite{polbos}
\bea
\partial_x\zeta^{>,<}(x,\tau) & = &\frac{1}{2\pi} \left[ p_+^{>,<}(x,\tau) - p_-^{>,<}(x,\tau) \right] \\
\partial_\tau \zeta^{>,<} (x,\tau) & = & -\frac{1}{2}\partial_x\zeta^{>,<} \left[ p_+^{>,<}(x,\tau) + p_-^{>,<}(x,\tau) \right]
\label{4-9}
\eea
These equations hold separately for each side of the potential, so that there are actually two collective fields.

We now determine the time evolution of the fermi surfaces using (\ref{4-8}) for the various quench profiles. Since the qualitative behavior of the solution is similar to abrupt quenches,we will use the latter expressions.

\begin{figure}[!h]
\centering
\includegraphics[width=4.0in]{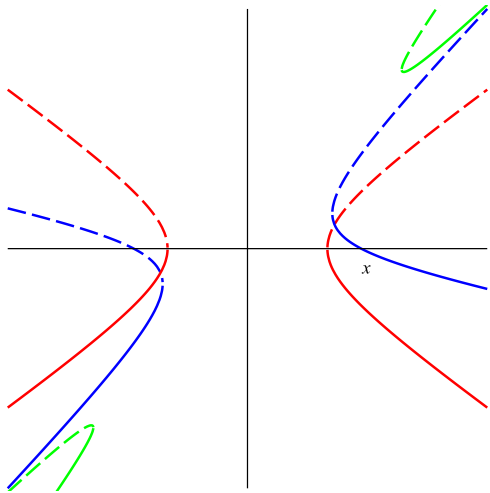}
\vspace{-2.1in}
\caption{The fermi surface profiles for $\omega_0 < \omega_1$ }
\label{fermisurface1}
\end{figure}

Figure \ref{fermisurface1} shows the profile of the fermi surfaces for the quench profile (\ref{abrupt}) for $\omega_0 < \omega_1$. The solid curves are the functions $p_-^{>,<}$ for different values of the time $\tau$ while the dashed curves are the functions $p_+^{>,<}$. These two curves meet at $x = \rho(\tau)$. The red curves are for $\tau=0$, and the blue and green curves are for later times. As time progresses the fermi surfaces fold onto themselves and receed from the potential on both sides.

\begin{figure}[!h]
\begin{centering}
\includegraphics[width=4.0in]{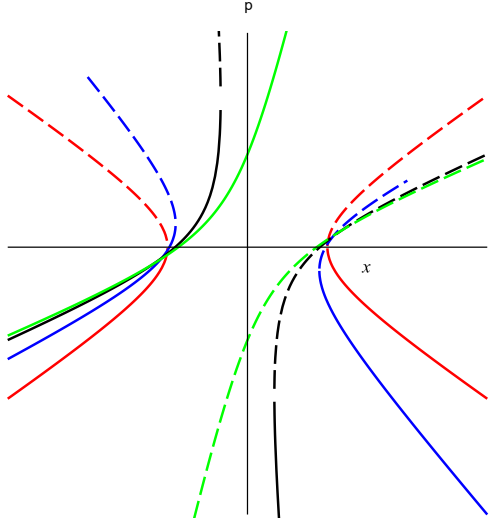}
\vspace{-2.1in}
\caption{The fermi surface profiles for $\omega_0 > \omega_1$ }
\label{fermisurface2}
\end{centering}
\end{figure}

Figure \ref{fermisurface2} shows the profile of the fermi surfaces for the quench profile (\ref{abrupt}) with $\omega_0 > \omega_1$. The solid curves are the functions $p_-^{>,<}$ for different values of the time $\tau$ while the dashed curves are the functions $p_+^{>,<}$. These two curves meet at $x = \rho(\tau)$. The red curves are for $\tau=0$, and the blue, black and green curves are for later times. In particular the black curves correspond to $\tau \sim \tau_0$ while the green curves are for $\tau > \tau_0$. It is clear that the fermi surfaces evolve smoothly across $\tau = \tau_0$, the time at which the collective description becomes problematic. As we approach $\tau = \tau_0$ the meeting place of these curves is pushed off to infinity. For $\tau > \tau_0$ the fermions are pushed to the other side, and only $p_-^<$ and $p_+^>$ are visible. As $\tau \rightarrow \infty$ two fermi surfaces $p_-^<$ and $p_+^>$ come close to each other, and it appears that the whole phase space is filled.

The smooth time evolution of the fermion theory now provides a meaning for the continuation of the saddle point collective field for $\tau > \tau_0$. We now have a single collective field which is defined by
\bea
\partial_x\zeta_0 & = &\frac{1}{2\pi}\left[ p_+^>(x,\tau) - p_-^<(x,\tau) \right] \\
& = & -\frac{1}{\trho(\tau)^2} \sqrt{x^2 + \trho(\tau)^2}
\label{4-10}
\eea
where $\trho$ is defined in (\ref{2-3bb}).

These profiles are for the double scaled potential. For finite $\beta$ (recall that $\beta \sim N$) the fermi surfaces are closed in phase space and the fermions cannot have arbitrarily large momenta and the entire phase space cannot be filled. In fact, for $\omega_0 > \omega_1$, the fermions which have large values of momenta as we approach $\tau = \tau_0$ come from the region of the original fermi sea which now lie above the ground state fermi level of the new potential. Consider e.g. a fermion which has $x=x_0$ and $p=p_0$ at $\tau = 0$, with $\omega_0^2 x_0^2 - p_0^2 > \omega_0^2$. If $\omega_1^2 x_0^2 - p_0^2 < \omega_1^2$ these fermions are above the ground state fermi surface of the new potential.  At $\tau = \tau_0 = \frac{1}{\omega_1}\tanh^{-1} (\frac{\omega_1}{\omega_0})$ the location and momentum of this fermion is
\ben
x(\tau_0) = \frac{\omega_0 x_0 + p_0}{\sqrt{\omega_0^2 - \omega_1^2}}
~~~~~~~
p(\tau_0) = \frac{\omega_1^2 x_0 + \omega_0 p_0}{\sqrt{\omega_0^2 - \omega_1^2}}
\label{4-12}
\een
Now consider a fermion which had a large negative $x_0$ and large positive $p_0$ at the time of the quench. Then $p_0 \approx - \omega_0 x_0$, and (\ref{4-12})  yields $x(\tau_0) \approx 0$ and $p(\tau_0) \approx -x_0 \sqrt{\omega_0^2 - \omega_1^2}$, which becomes arbitrarily large when $x_0$ becomes large. However, when $\beta$ is finite one needs to modify the inverted harmonic potential at large values of $x$ by e.g. imposing a hard wall at $|x| \sim \sqrt{\beta}$. This means that the maximum momenta are also of the order $\sqrt{\beta}$.

\begin{figure}
\centering
\subfloat[$\tau=0$]{
\includegraphics[width=0.4\textwidth]{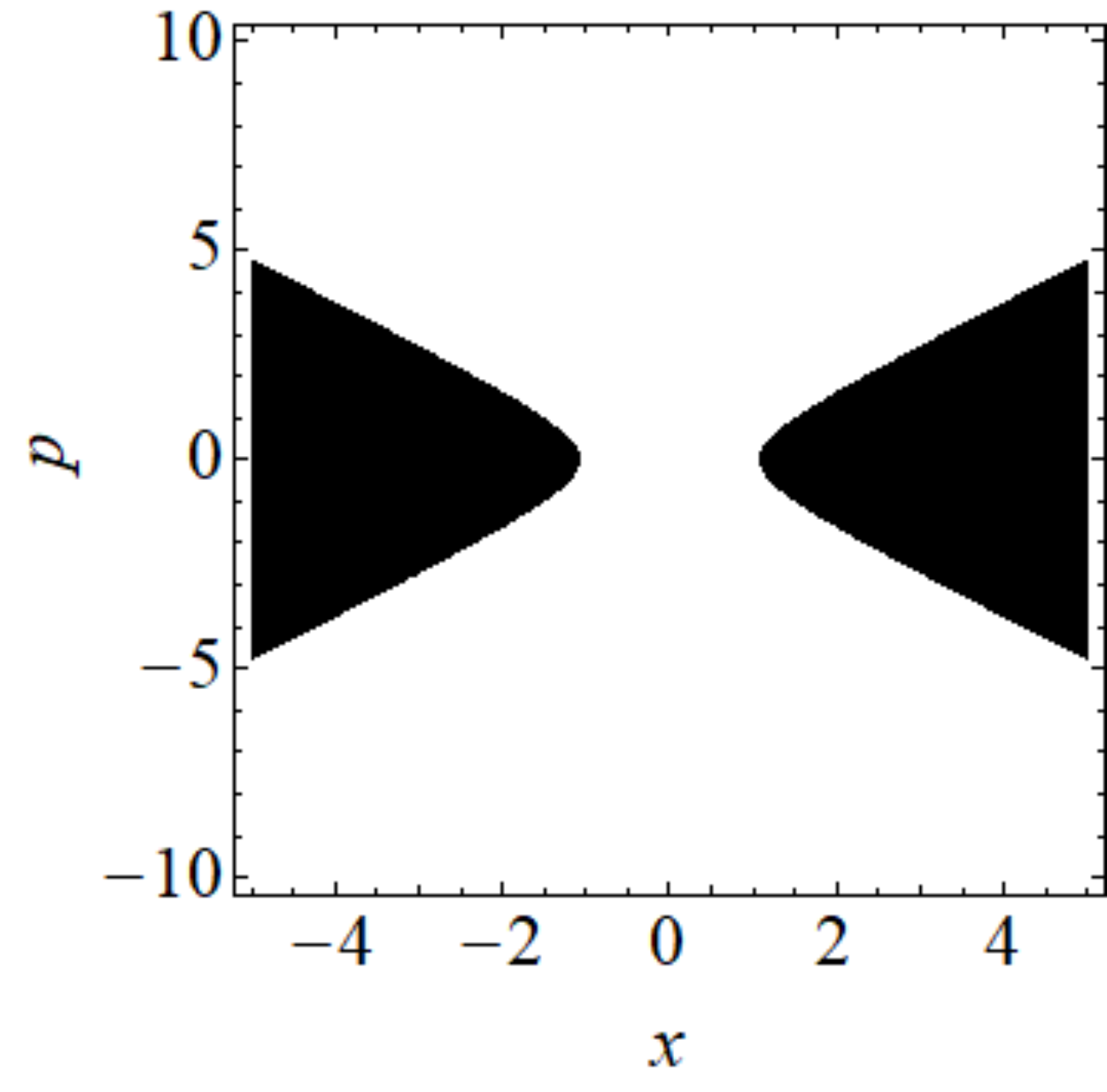}}
\subfloat[$\tau=\frac{1}{8}$]{
\includegraphics[width=0.4\textwidth]{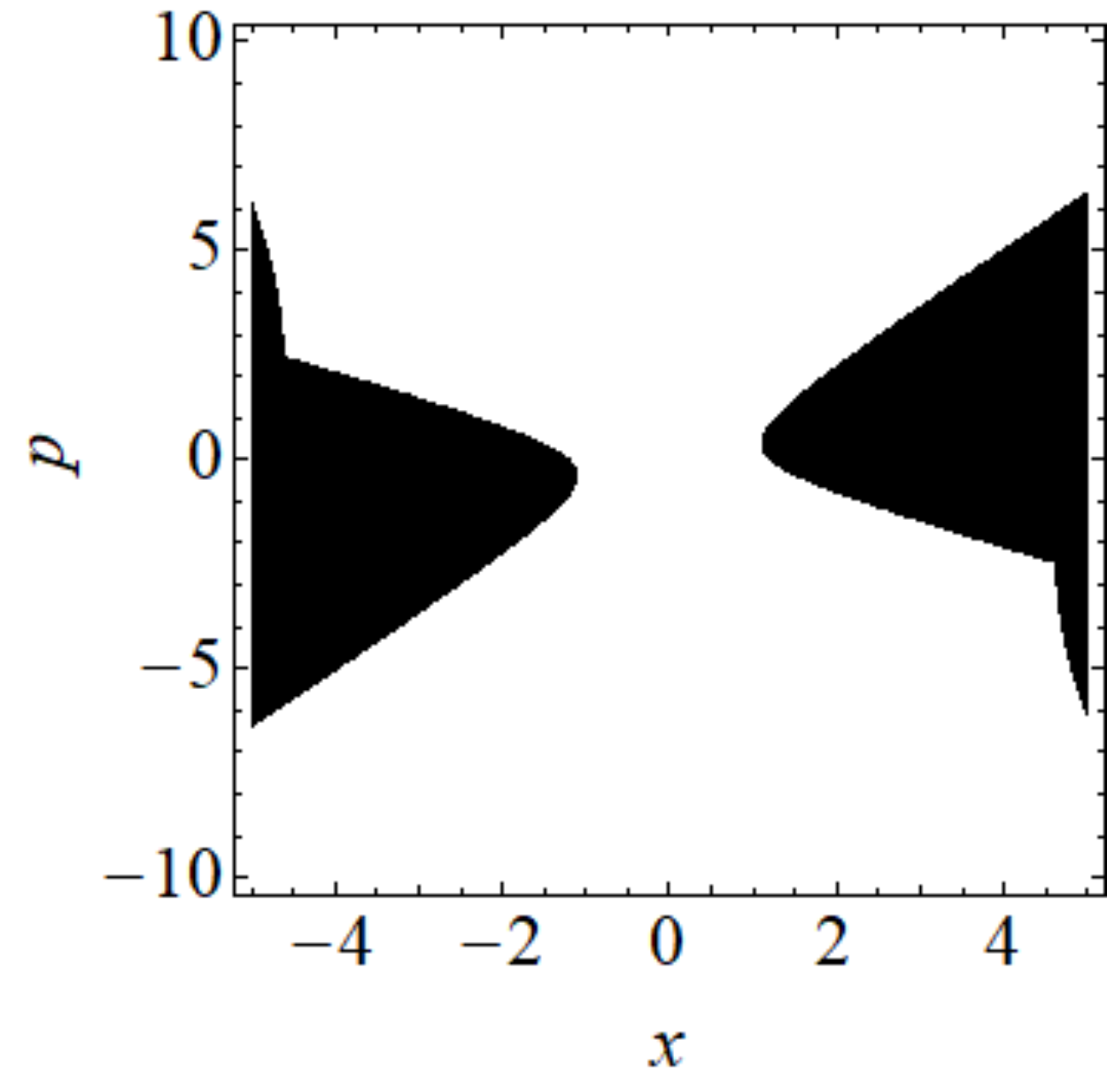}}\\
\subfloat[$\tau=\frac{1}{4}$]{
\includegraphics[width=0.4\textwidth]{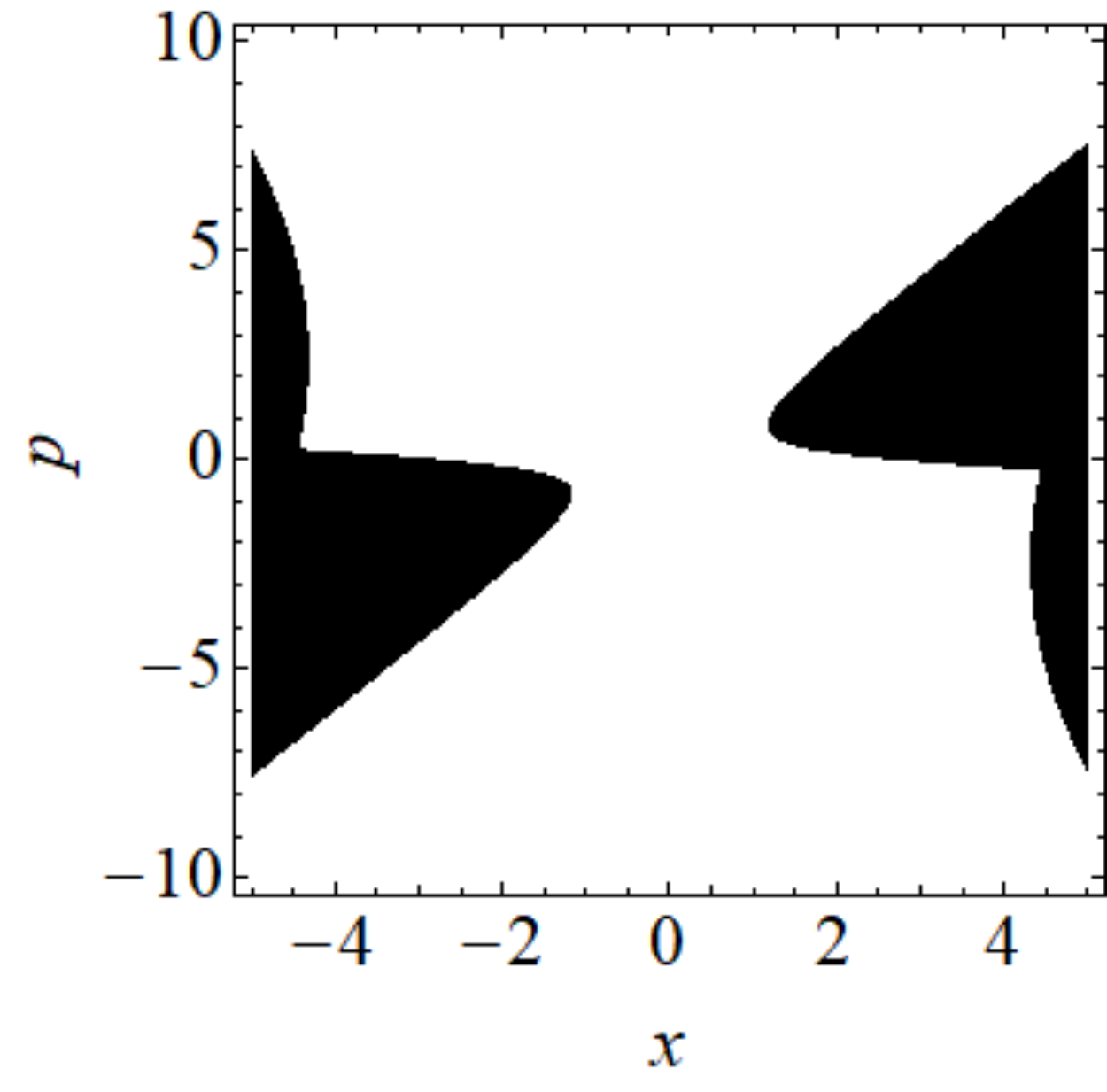}}
\subfloat[$\tau=\frac{1}{2}$]{
\includegraphics[width=0.4\textwidth]{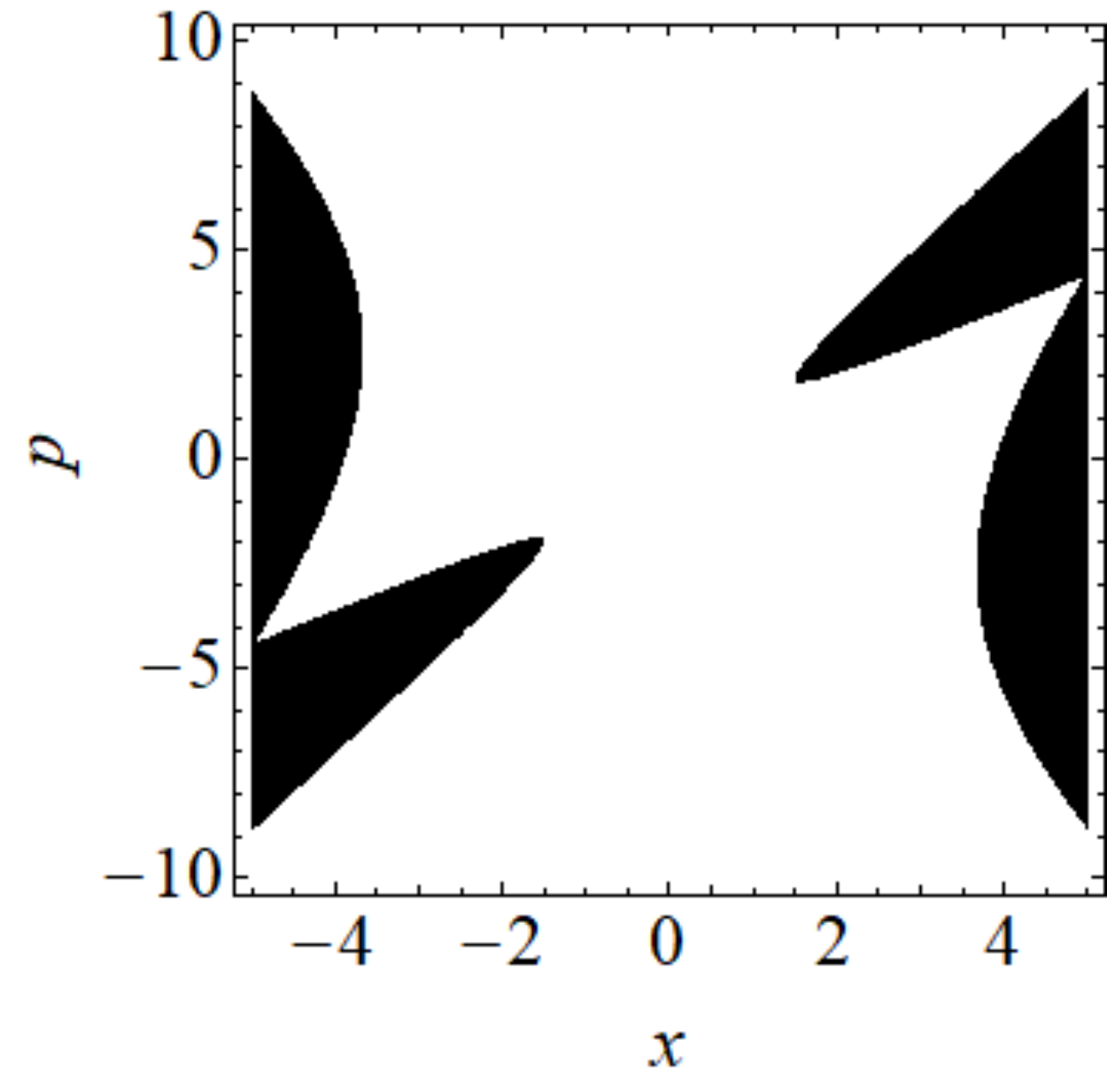}}\\
\caption{Time evolution of phase space density $u(x, p, \tau)$ for the potential with a cutoff at $x=\pm l/2$ after abrupt quench. Choose $\omega_0 = \frac{1}{2},\omega_1 =1$, $l/2=5$.}
\label{psdl}
\end{figure}

\begin{figure}
\centering
\subfloat[$\tau=0$]{
\includegraphics[width=0.4\textwidth]{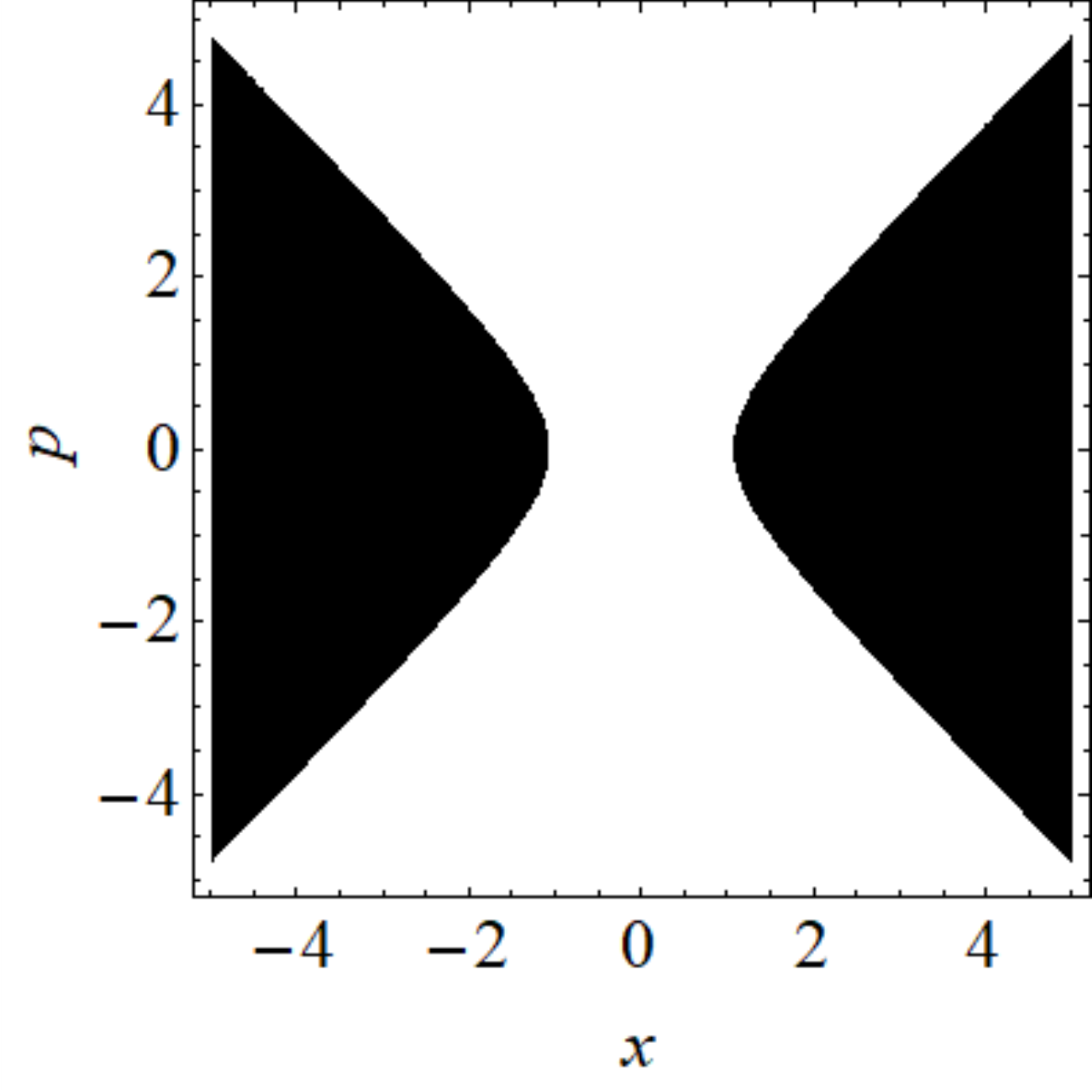}}
\subfloat[$\tau=\frac{1}{2}\log 3-0.0001 $]{
\includegraphics[width=0.4\textwidth]{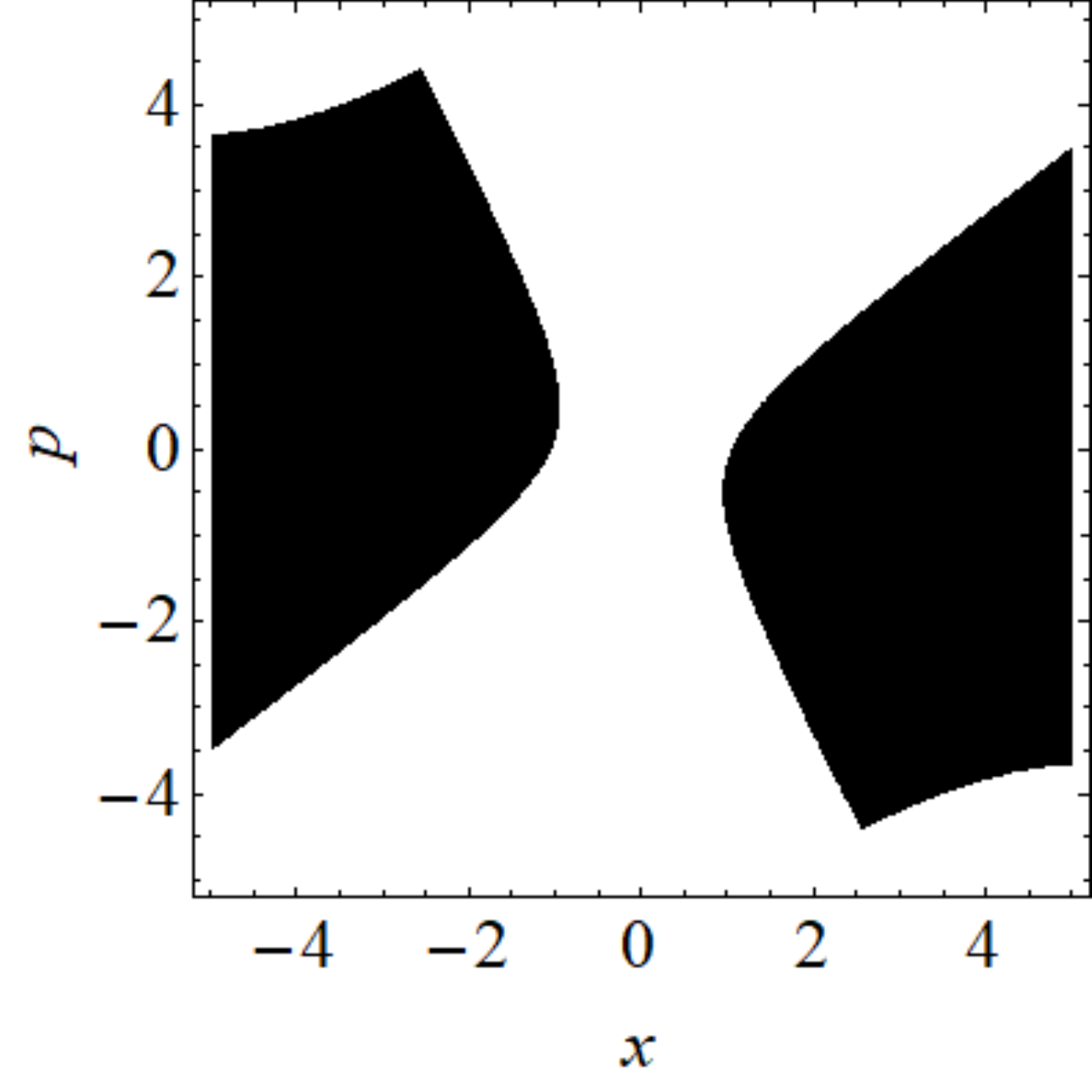}}\\
\subfloat[$\tau=\log 3-0.001$]{
\includegraphics[width=0.4\textwidth]{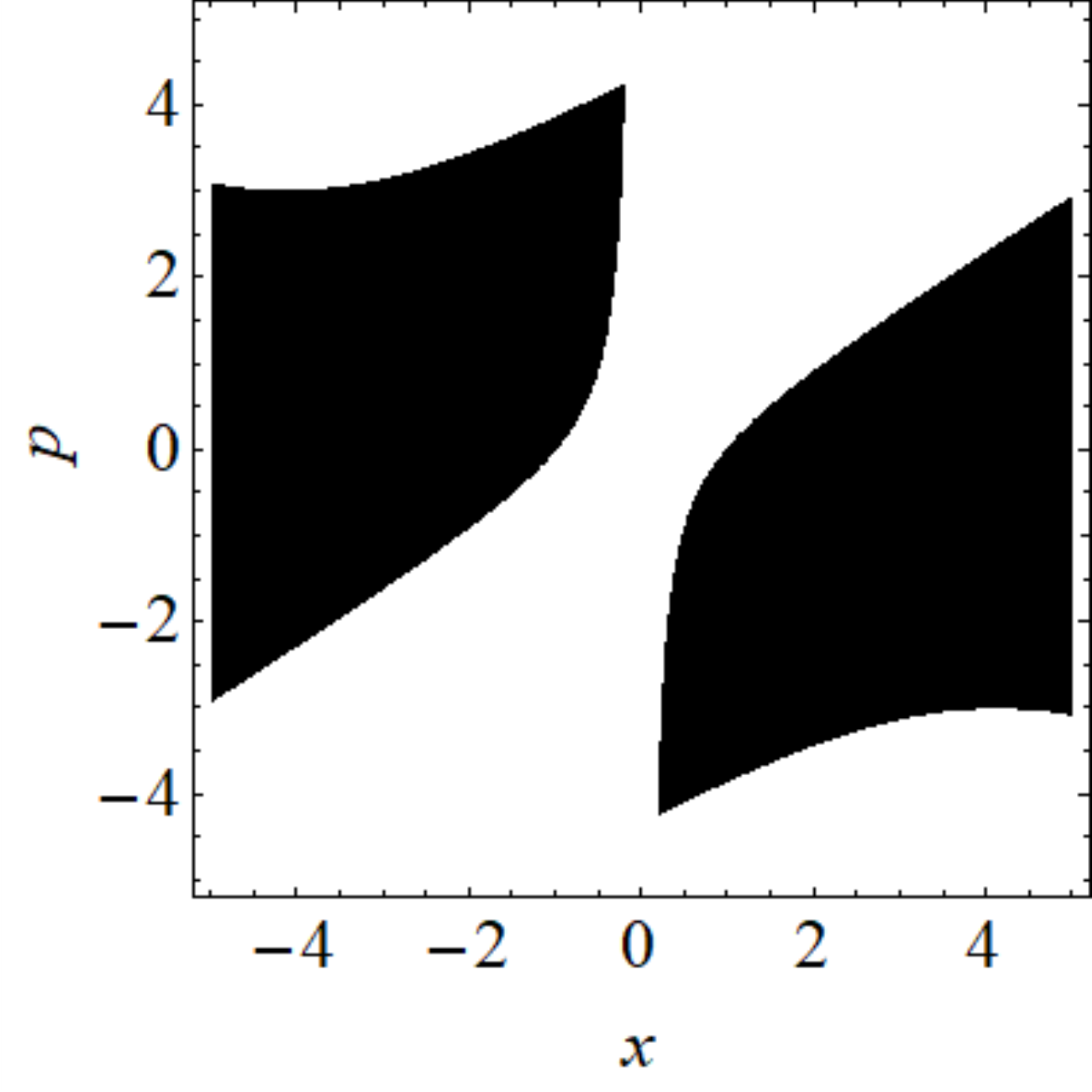}}
\subfloat[$\tau= \frac{3}{2}\log 3 -0.0001$]{
\includegraphics[width=0.4\textwidth]{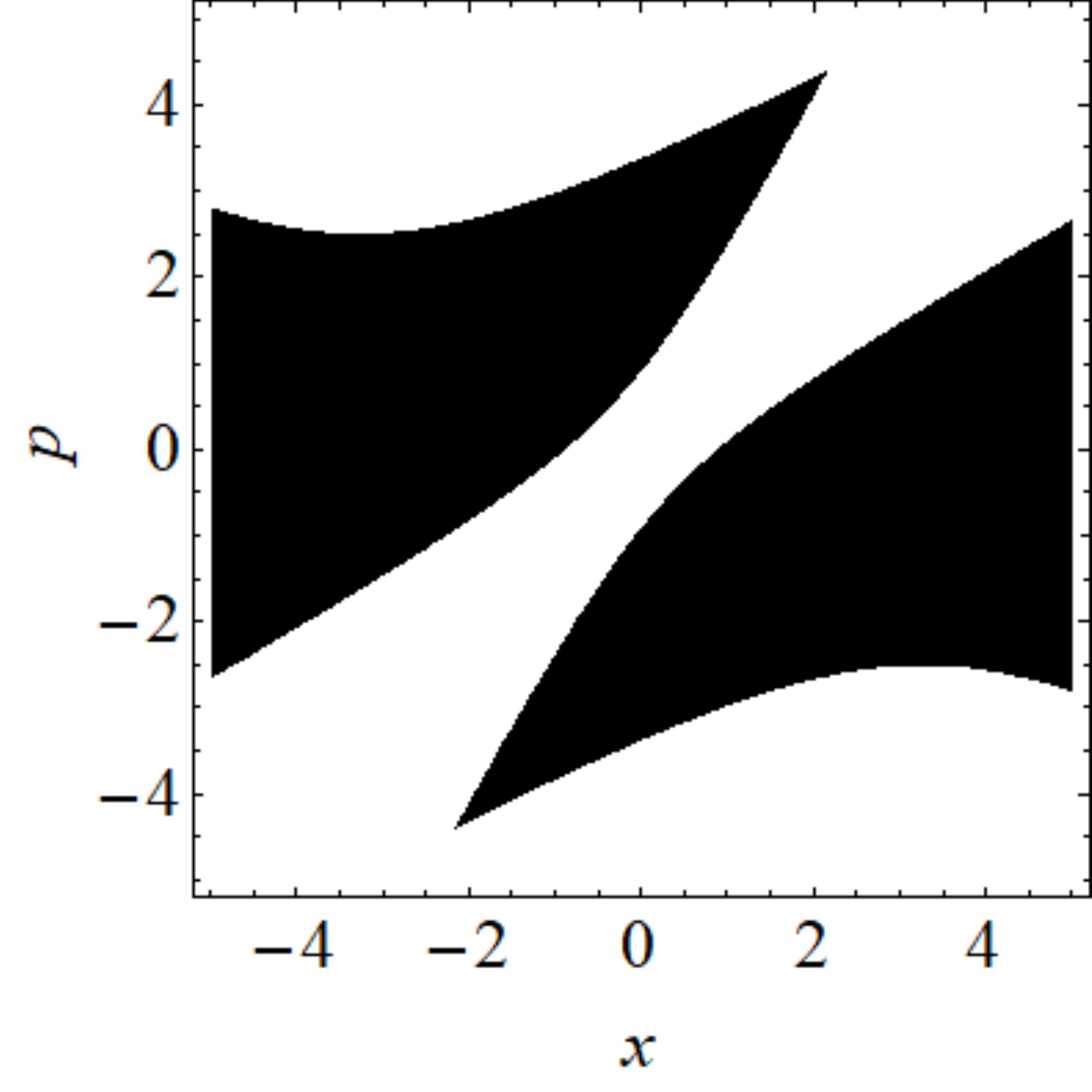}}\\
\caption{Time evolution of phase space density $u(x, p, \tau)$ for the potential with a cutoff at $x=\pm l/2$ after abrupt quench. Choose $\omega_0 = 2\omega_1 =1$, $l/2=5$.}
\label{psds}
\end{figure}

The fermi surface profiles for a potential with a cutoff are shown in figure \ref{psds} and figure \ref{psdl} for various times.  Here the filled regions of the phase space are shown in black. At an early time after abrupt quench, each fermi surface profile includes two parts: part of the fermi surface profile for potential without cutoff (shown in figure \ref{fermisurface1} and \ref{fermisurface2}), and its reflection after hitting the cutoff. In figure \ref{fermisurface1}, particles flow out of the imaginary cutoff while no particles flow in, thus the reflection fermi surface profile should make up the loss to keep the fermion number (the area of black region in figure \ref{psdl}) invariant. However, in figure \ref{fermisurface2}, more particles with positive energy flow in than flow out, thus the filled fermi surface in the presence of a cutoff excludes these fermions. We explain the details of these two different cases and give the expression of the phase space density $u(x,p, \tau)$ in appendix \ref{apppsd}.

\section{The emergent space-time} 

The nature of the emergent space-time is deduced from the fluctuation action. In particular the quadratic action for fluctuations (\ref{2-11}) represents a massless scalar field in a $1+1$ dimensional space-time. Since a massless scalar field is insensitive to a conformal factor, we can only determine the conformal class of the metric. Equation (\ref{2-13}) is one member of this class. The non-trivial features of the emergent space-time are its global aspects - this is what we need to determine.

For this purpose it is useful, as always, to find coordinates such that the metric becomes conformal to standard Minkowskian metric, which is always possible in $1+1$ dimensions. For an arbitrary quench profile, the Minkowskian coordinates $(q,T)$ are given by,
\ben
x = \pm \trho(\tau) \cosh (q) ~~~~~~T = \int^\tau \frac{d\tau'}{\rho(\tau)^2}
\label{6-1}
\een
This is adequate for the single step abrupt quench with $\omega_0 < \omega_1$. In this case, we can explicitly obtain, for $\tau > 0$
\bea
x & = &\pm\frac{1}{\sqrt{\omega_0}}\left[\cosh^2\omega_1 \tau - \frac{\omega_0^2}{\omega_1^2} \sinh^2\omega_1 \tau \right]^{1/2} \cosh q \\
T & = & \tanh^{-1}\left[ \frac{\omega_0}{\omega_1}\tanh (\omega_1 \tau) \right]
\label{6-2}
\eea
Note that the forbidden values of $x$ increase with time.
As $0 < \tau < \infty$ we have $ 0 < T < \tanh^{-1} \frac{\omega_0}{\omega_1}$. The space-time appears to be geodesically incomplete, and normally one would have extended the time $T$ further to $\infty$. However the underlying matrix model tells us that this extension does not make any sense, since the matrix model time $\tau$ ends at this point. In fact, on the space-like line $T = \tanh^{-1} \frac{\omega_0}{\omega_1}$, the cubic couplings of the fluctuation field diverge - in this sense this is a space-like singularity. As mentioned in the introduction this means that the semi-classical collective theory as well as the fermi fluid description are not valid anymore and strong coupling effects become important. The situation is similar to such singularities in GR which signal the breakdown of Einstein's equations.

\begin{figure}
\centering
\includegraphics[width=0.5\textwidth]{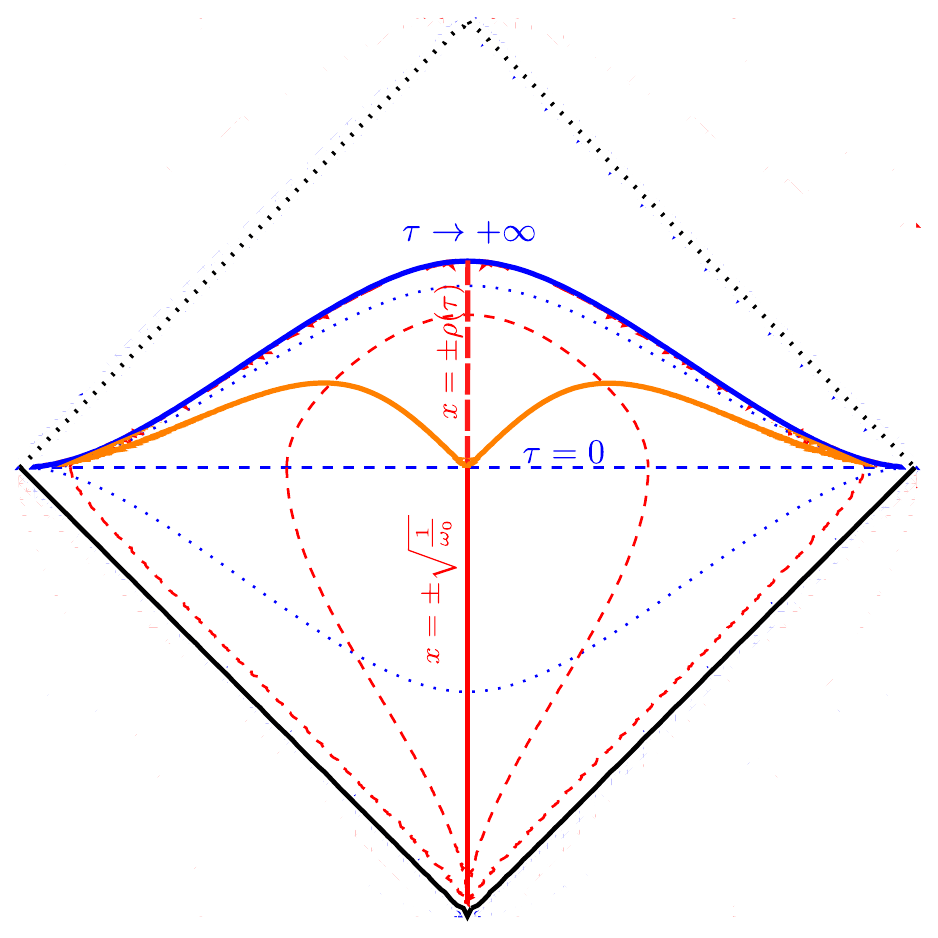}
\caption{Penrose diagram for emergent space-time when $\omega_0 =\frac{1}{\sqrt{2}} \omega_1=1$. Blue dotted lines are constant $\tau$ lines; Specially, $\tau=0$ when the abrupt quench occurs are plotted in blue dashed line; $\tau \to \infty$ i.e. infinite future is plotted in blue solid line. Red dashed lines are constant $x$ lines. Special values of $x$, $x=\pm \sqrt{\frac{1}{\omega_0}}$ and $x=\pm \rho(\tau)$, is plotted in thick red solid line and thick red dashed line, respectively. The two sides of these lines are different disconnected space-times where the fluctuations of the left and right fermi surface propagate. The orange solid lines demarcate the regions in which constant $x$ lines are spacelike from those where they are time-like.}
\label{penrose1}
\end{figure}

An interesting feature of this space-time is that the $x = {\rm constant}$ lines do not remain time-like for all times. The Penrose diagram for the emergent space-time is shown in figure \ref{penrose1}. In this figure the thick red solid line and the thick red dashed line seperate two different pieces of the space-time which are perceived by the fluctuations of the left and right fermi surfaces. 
The red dashed lines are constant $x$ lines, while the blue dotted lines are constant $\tau$ lines. Clearly the constant $\tau$ lines are always spacelike. This can be seen from the form of the metric (\ref{2-13}) since $ds^2 > 0$ for $d\tau = 0$. However constant $x$ lines are timelike only when
\ben
\frac{(\partial_\tau \zeta_0)^2}{\pi^2 (\partial_x \zeta_0)^4} \leq 1
\label{6-3}
\een
They are null when the equality is satisfied and space-like otherwise.


\begin{figure}
\centering
\includegraphics[width=0.5\textwidth]{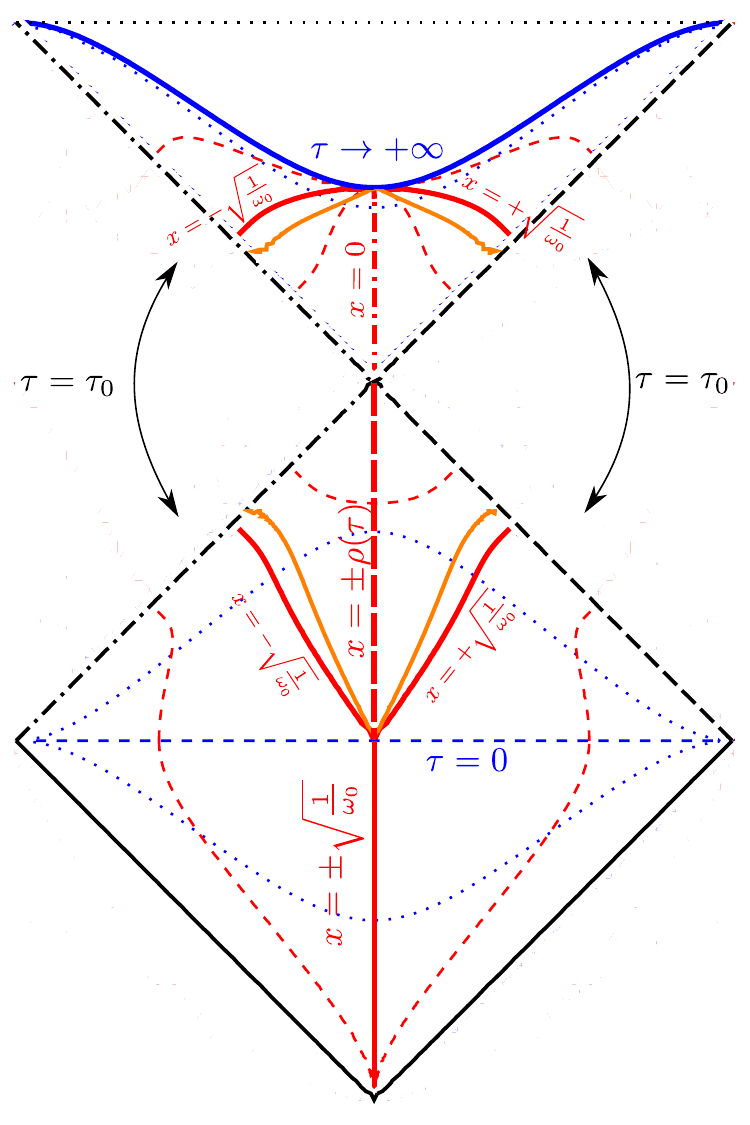}
\caption{Penrose diagram for emergent space-time when $\omega_0 =\sqrt{2} \omega_1=1$. The black dot-dashed lines and dashed lines represent $\tau = \tau_0$ where $\tau_0$ is defined in (\ref{3-3}) and should be glued respectively. Blue dotted lines are constant $\tau$ lines; Specially, $\tau=0$ when the abrupt quench occurs are plotted in the blue dashed line; $\tau \to \infty$ i.e. infinite future is plotted in blue solid line. Red dashed lines are constant $x$ lines. The special value of $x$, $x=\pm \sqrt{\frac{1}{\omega_0}}$, is plotted in thick red solid lines; it splits into two lines at $\tau =0$. $x=\pm \rho(\tau)$, is plotted in thick red dashed line. These separate two disconnected space-times corresponding to the fluctuations of the left and right fermi surfaces. These two pieces connect at $\tau=\tau_0$. $x=0$ is plotted in the thick red dot-dashed line - this, however does not separate disconnected pieces.
The orange solid lines demarcate the regions in which constant $x$ lines are spacelike from those where they are time-like.}
\label{penrose2}
\end{figure}

For an abrupt quench with $\omega_0 > \omega_1$, the coordinates $(T,q)$ defined in (\ref{6-2}) cover only the region $\tau < \tau_0$, and the line $\tau = \tau_0$ has $T = \infty$, and is null. The red solid line is the line $x= \pm \frac{1}{\sqrt{\omega_0}}$. For $\tau \leq 0$ this is $q=0$ which separates the two sides of the potential. For $\tau > 0$ excitations can propagate in the region $|x| < \frac{1}{\sqrt{\omega_0}}$, so this region is included in the Penrose diagram. The orange line demarcates the regions where constant $x$ surfaces (denoted by red dashed lines) are space-like and time-like. Finally these constant $x$ lines become null as they approach the $\tau = \tau_0$ lines. The part of the Penrose diagram for $\tau < \tau_0$ actually consists of two disconnected pieces (corresponding to the fluctuations of the left and right fermi surfaces) separated by the red solid line for $\tau < 0$ and the red dashed line for $\tau >0$.

Normally $T = \infty$ or $\tau= \tau_0$ would be the future null boundary of two dimensional Minkowski space and there is no reason for continuing the space-time beyond this. However the matrix model predicts a smooth time evolution beyond $\tau = \tau_0$ - we therefore need to attach another piece of space-time. The Minkowskian coordinates are now defined by 
\ben
x =  \trho (\tau) \sinh (q)~~~~~~T =- \coth^{-1} \left[ \frac{\omega_0}{\omega_1} \tanh (\omega_1\tau) \right]
\label{6-4}
\een
In the Penrose diagram this piece is given by the upper triangle. Note that the two disconnected pieces for $\tau < \tau_0$ have joined into a single connected space-time.
Equation (\ref{6-4}) shows that as $\tau$ ranges from $\tau_0$ to $\infty$ the time $T$ ranges from $-\infty$ to a finite value $\coth^{-1}\left( \frac{\omega_0}{\omega_1} \right)$. The matrix model tells us to end the space-time here. The cubic couplings diverge here - so this is like a space-like singularity. Again this means that our semi-classical treatment, together with the space-time interpretation of the model, breaks down here. The matrix model time evolution is possibly smooth, but this requires an exact treatment which we have not pursued. 

Once again, the constant $x$ lines are not always timelike. The resulting Penrose diagram is shown in figure \ref{penrose2}. 

For the finely tuned case, the space-time is geodesically complete : at late times $\rho^2$ saturates so that as $\tau \rightarrow \infty$ one also has $T \rightarrow \infty$.

\section{Conclusions} In this work we explored the nature of emergent space-times in the $c=1$ matrix model with a time dependent coupling. Our main finding is that generically such a quantum quench naturally leads to geodesically incomplete space-times with space-like boundaries where the coupling of the "bulk" theory diverges. Only for very finely tuned quenches, the emergent space-time is normal. 

It will be interesting to understand the time dependence of the entanglement entropy of a region of the emergent space-time. In \cite{dhl} it was shown that in a system of free fermions in an external potential, the entanglement entropy $S_A$ of a region of the eigenvalue space $A$ in the "in" state $|in \rangle$ (i.e. the Heisenberg picture state which is the ground state of the initial hamiltonian) can be expressed entirely in terms of the phase space density $u(x,p,\tau)$,
\bea
S_A & = & {1\over2\pi}\int_{-\infty}^\infty dp  \int_A dx \, \langle in|u(p,x,\tau)|in\rangle - \nn \\
& & {1\over(2\pi)^2}\int_{-\infty}^\infty dp_1 dp_2 \int_A dx dy~ e^{-i(p_2-p_1)(x-y)}~\langle in|u(p_1,(x+y)/2,\tau)|in\rangle \langle in|u(p_2,(x+y)/2,\tau)|in\rangle\nn\\
\label{7-1}
\eea
The solutions for the phase space density for the various quench profiles studied in this paper can be then used to compute this quantity. For constant frequencies this bulk entanglement entropy has been investigated in \cite{entang} : one key result of this investigation is that this quantity is finite, with the string coupling replacing the usual UV cutoff. However the meaning of this quantity for our time dependent bulk space-time is unclear at the moment, since constant $x$ lines do not remain time-like for all times. We are currently investigating this issue.

From the point of view of holography, the most interesting question is the relation of these to solutions to those of two dimensional string theory in the worldsheet or string field theory formulations. For the standard time independent case, this relationship 
involves a transformation whose kernel is non-local at the string scale, and obtained by a comparison on the S matrices obtained from the matrix model and worldsheet string theory. This is a manifestation of the "leg-pole" factors. One may be able to obtain a worldsheet description using the connection of the feynman diagrams of the matrix model with dynamical triangulations of the worldsheet. Naively this would lead to a time dependent dilaton and a time dependent worldsheet cosmological constant. However a precise relationship could be subtle and is under investigation. A precise connection to a worldsheet S-matrix can be then established. 
Assuming, however, that the non-locality continues to remain at the string scale, the space-time diagrams discussed above need to be smeared at this scale. As mentioned above, the single step abrupt quench results are identical to the time dependent solutions of the constant coupling matrix model considered in \cite{dk} : this work contains some discussions of this issue.

 \section*{Acknowledgements}
S.R.D. would like to thank Gautam Mandal for many discussions and clarifications. 
Some of the results of this work were presented in talks given by S.R.D. at the workshops "Quantum Information and String Theory 2019" at Yukawa Institute, Kyoto and "Singularities and Horizons: From Black Holes to Cosmology" at Lorentz Center, Leiden, S.R.D. would like to thank the oragnizers and participants of these workshops for discussions and hospitality. S.R.D. would also like to thank the Tata Institute of Fundamental Research and Brown University for hospitality during the completion of this work, and the audience of seminars given there for questions which clarified many aspects. The work of S.R.D and S.L are partially supported by National Science Foundation grant NSF/PHY-1818878. The work of S.H is supported by the Lyman T. Johnson postdoctoral fellowship.

\begin{appendix}
\section{Abrupt Pulse and Dip protocols $\omega_0\to\omega_2\to\omega_1$}\label{abrupt ccp}
Here we compute analytic solutions for $\rho(\t)^2$ for an abrupt pulse with $\omega_0\to\omega_2\to\omega_1$ with an $f(\t)^2$ of
\bea
f(\t)^2 
=\begin{cases} \omega_0^2,  \qquad \t< -{T\over2}\\
\omega_2^2,  \qquad  -{T\over2}\leq \t < {T\over2}\\
 \omega_1^2,  \qquad    {T\over2}\leq \t 
\end{cases}
\label{f}
\eea
We know that the solution to the Pinney equation is given by 
\bea
\rho^2(\t) = A u(\t)^2 + 2B u(\t)v(\t) + Cv(\t)^2
\eea
where $u(\t)$ and $v(\t)$ are independent solutions which satisfy
\bea
{d^2u\over d\tau^2 } -f(\t)^2 u &=&0\cr
{d^2v\over d\tau^2 } -f(\t)^2 v &=&0
\eea
We require for $\t < - {T\over2}$ that $\rho^2 = {1\over\omega_0}$. This requires $A = C = 0$ and by (\ref{2-9}), $2B={1\over\omega_0}$. Therefore
\bea
\rho^2 =  {1\over\omega_0} u(\t)v(\t) 
\label{rho sq a}
\eea
We look for solutions
\bea
{d^2u\over d\tau^2 } -f(\t)^2 u &=&0\cr
{d^2v\over d\tau^2 } -f(\t)^2 v &=&0
\eea
for $f(\t)^2$ given in (\ref{f}). A general solution for $u(\t)$ in the three regions is given by
\bea
u_1(\t) &=& e^{\omega_0\t}\cr
u_2(\t) &=& Ae^{\omega_2\t} +  Be^{-\omega_2\t}\cr
u_3(\t) &=& Ce^{\omega_1\t} +  De^{-\omega_1\t}
\label{ua}
\eea
Enforcing boundary conditions at $\t=-{T\over2}$ and $\t={T\over 2}$, we find
\bea
 e^{-\omega_0{T\over2}} &=& A e^{-\omega_2{T\over2}} + B e^{\omega_2{T\over2}} \cr
 \omega_0 e^{-\omega_0{T\over2}}&=& \omega_2 ( A e^{-\omega_2{T\over2}} - B e^{\omega_2{T\over2}}  )\cr
  A e^{\omega_2{T\over2}} + B e^{-\omega_2{T\over2}}  &=&   C e^{\omega_1{T\over2}} + D e^{-\omega_1{T\over2}}\cr
\omega_2 ( A e^{\omega_2{T\over2}} - B e^{-\omega_2{T\over2}}  ) &=&   \omega_1 ( C e^{-\omega_1{T\over2}} - D e^{\omega_1{T\over2}}  )
\eea
Solving for the coefficients yield
\bea
A &=&e^{{1\over2}T(-\omega_0 + \omega_2)}{(\omega_0 + \omega_2)\over2\omega_2} \cr
B&=& e^{-{1\over2}T(\omega_0 + \omega_2)}{(-\omega_0 + \omega_2)\over2\omega_2}\cr
C&=& e^{-{T\over2}(\omega_0 + \omega_1+2\omega_2)}{((\omega_0 - \omega_2)(-\omega_1 + \omega_2) + e^{2T\omega_2}(\omega_0 +\omega_2)(\omega_1+\omega_2))\over 4\omega_1\omega_2}\cr
D&=&e^{-{T\over2}(\omega_0 - \omega_1+2\omega_2)}{( e^{2T\omega_2}(\omega_0 + \omega_2)(\omega_1 - \omega_2) -(\omega_0 -\omega_2)(\omega_1+\omega_2))\over 4\omega_1\omega_2}
\label{u coeff a}
\eea
Similarly, for the other independent solution $v(\t)$ we have
\bea
v_1(\t) &=& e^{-\omega_0\t}\cr
v_2(\t) &=& A'e^{\omega_2\t} +  B'e^{-\omega_2\t}\cr
v_3(\t) &=& C'e^{\omega_1\t} +  D'e^{-\omega_1\t}
\label{va}
\eea
and again applying boundary conditions 
\bea
 e^{\omega_0{T\over2}} &=& A' e^{-\omega_2{T\over2}} + B' e^{\omega_2{T\over2}} \cr
- \omega_0 e^{\omega_0{T\over2}}&=& \omega_2 ( A' e^{-\omega_2{T\over2}} - B' e^{\omega_2{T\over2}}  )\cr
  A' e^{\omega_2{T\over2}} + B' e^{-\omega_2{T\over2}}  &=&   C' e^{\omega_1{T\over2}} + D' e^{-\omega_1{T\over2}}\cr
\omega_2 ( A' e^{\omega_2{T\over2}} - B' e^{-\omega_2{T\over2}}  ) &=&   \omega_1 ( C' e^{-\omega_1{T\over2}} - D' e^{\omega_1{T\over2}}  )
\eea
we find
\bea
A' &=& e^{{1\over2}T(\omega_0 + \omega_2)}{(-\omega_0 + \omega_2)\over2\omega_2} \cr
B'&=& e^{{1\over2}T(\omega_0 - \omega_2)}{(\omega_0 + \omega_2)\over2\omega_2}\cr
C'&=& e^{{T\over2}(\omega_0 - \omega_1 - 2\omega_2)}{((\omega_0 + \omega_2)(\omega_1 - \omega_2) + e^{2T\omega_2}(-\omega_0 +\omega_2)(\omega_1+\omega_2))\over 4\omega_1\omega_2}\cr
D'&=&e^{{T\over2}(\omega_0 + \omega_1 - 2\omega_2)}{( e^{2T\omega_2}(\omega_0 - \omega_2)(-\omega_1 + \omega_2)  + (\omega_0 + \omega_2)(\omega_1+\omega_2))\over 4\omega_1\omega_2}
\label{v coeff a}
\eea
Therefore, inserting (\ref{ua}) and (\ref{va}) into (\ref{rho sq a}) with the coefficients defined in (\ref{u coeff a}) and (\ref{v coeff a}), $\rho^2(\t)$ becomes
\bea
\rho^2(\t)&=&\begin{cases}
 {1\over\omega_0},\qquad \t \leq -{T\over2}\\
\\
{1\over\omega_0}\big(AA'e^{2\omega_2\t}+BB'e^{-2\omega_2\t} + AB'+A'B\big),\qquad  -{T\over2}\leq\t \leq {T\over2}
\\
\\
{1\over\omega_0}\big(CC'e^{2\omega_1\t}+DD'e^{-2\omega_1\t} + CD'+C'D\big),\qquad \t \geq  {T\over2}
\end{cases}
\label{rhos}
\eea

\section{Smooth step protocols $\omega_0\to\omega_1$}\label{smooth ecp}
Here we compute analytic solutions for $\rho(\t)^2$ for a smooth step for $\omega_0\to\omega_1$ with an $f(\t)^2$ of
\bea
f(\t)^2  &=&\omega_0^2+ {\omega_1^2 -\omega_0^2 \over 1 + e^{-{\t\over \d t}} }\cr
&=& \frac{\omega_1^2 + \omega_0^2 e^{-\frac{\tau}{\dt}}}{1+ e^{-\frac{\tau}{\dt}}}
\label{smooth ECP profile}
\eea
where in this case we choose
\bea
\rho^2(\t) = u(\t)v(\t) 
\label{rho sq 2}
\eea
First we look for solutions to
\bea
{d^2u'\over d\tau^2 } -f(\t)^2 u' &=&0\cr
{d^2v'\over d\tau^2 } -f(\t)^2 v' &=&0
\eea
where in general our solution will be of the form
\bea
u(\t)&=&A' u'(\t) + B' v'(\t)\cr
v(\t)&=&A u'(\t) + B v'(\t)
\eea
We find that the two independent solutions are given by
\bea
u'(\t) &=&e^{\omega_0 \t}{}_2F_1 (\d t (\omega_0 - \omega_1), \d t( \omega_0 + \omega_1 ), 1 + 2\delta t\omega_0, -e^{{\t\over\delta t}}) \cr
v'(\t)&=& e^{-\omega_0 \t}{}_2F_1 (-\d t (\omega_0 +\omega_1), \d t(-\omega_0 + \omega_1 ) , 1 - 2\delta t\omega_0, -e^{{\t\over\delta t}})
\label{u' and v'}
\eea
The $v'$ solution is nonsingular for $\omega_0\d t$ not a half integer. 

\subsection{Solution for $u$}
Now we write the following adiabatic solution for $u$.
\bea
u_{ad}(\t) &=&{1\over\sqrt{\omega(\t)}} e^{-C(\d t,\omega_0,\omega_1)+\int^{\t}f(\t')d\t'}\cr
&=&  {1\over\sqrt{\omega(\t)}}e^{-C(\d t,\omega_0,\omega_1)+\int^{\t} \big(\omega_0^2 + {\omega_1^2-\omega_0^2\over 1 + e^{-{\t'\over\d t}}}\big)^{1\over2}d\t'}\cr
&=&{1\over\sqrt{\omega(\t)}} e^{-C(\d t,\omega_0,\omega_1)+F(\t,\d t,\omega_0,\omega_1)}
\label{adia u}
\eea
where
\bea
C(\d t,\omega_0,\omega_1) &=&-\d t\omega_0\log(4\omega_0^2) + \d t\omega_1\log(\omega_0 +\omega_1)^2\cr
F(\t,\d t,\omega_0,\omega_1) &=&   \bigg(\omega _0 \left(\t-\text{$\delta $t} \log \left(2 \omega _0
   \sqrt{e^{\t/\text{$\delta $t}}+1} \sqrt{\omega _1^2 e^{\t/\text{$\delta $t}}+\omega _0^2}+\left(\omega
   _0^2+\omega _1^2\right) e^{\t/\text{$\delta $t}}+2 \omega _0^2\right)\right)\cr
   &&\quad+\text{$\delta $t} \omega _1
   \log \left(\omega _1 \left(2 \omega _1 e^{\t/\text{$\delta $t}}+2 \sqrt{e^{\t/\text{$\delta $t}}+1}
   \sqrt{\omega _1^2 e^{\t/\text{$\delta $t}}+\omega _0^2}+\omega _1\right)+\omega
   _0^2\right)\bigg)\nn \\
   \label{coeff}
\eea
The constant $C(\d t,\omega_0,\omega_1)$ is chosen such that for $\t \to - \infty$
\bea
u_{ad}\to {1\over\sqrt{\omega_0}}e^{\omega_0\t}
\eea
Therefore, the coefficient $C_u$ in (\ref{5-3}), is given by
\bea
C_u=e^{-C(\d t,\omega_0,\omega_1)}
\eea
 The derivative of $u_{ad}$ is given by
\bea
\partial_{\t} u_{ad}(\t) &=& \bigg(f(\t) - {\partial_{\t}f(\t)\over2f(\t)} \bigg) u_{ad}(\t) 
\label{der adia u}
\eea
Solving for the solution
\bea
u(\t) = A'u'(\t) + B' v'(\t)
\label{u}
\eea
we impose the boundary conditions
\bea
u(T) &=& u_{ad}(T)\cr
\partial_{\t} u(T)&=& \partial_{\t}u_{ad}(T)
\eea
This yields an $A'$ and $B'$ of
\bea
A' &=& {1\over 4}e^{-T\omega_0-C(\d t,\omega_0,\omega_1)+F(T,\d t,\omega_0,\omega_1) }\bigg(\omega_1^2 + {\omega_0^2 - \omega_1^2\over 1 + e^{{T\over\d t}}}\bigg)^{3\over4}\cr
&&\bigg[\bigg(4\d t \omega_0^2\bigg(-\omega_1 +\sqrt{\omega_1^2 + {\omega_0^2 - \omega_1^2\over 1 + e^{{T\over\d t}}}} \bigg) + 4e^{2{T\over\d t}}\d t \omega_1^2\bigg(-\omega_1 +\sqrt{\omega_1^2 + {\omega_0^2 - \omega_1^2\over 1 + e^{{T\over\d t}}}} \bigg)\cr
&&\qquad + e^{T\over\d t}\omega_0^2\bigg(1 + 4\d t\bigg(-\omega_1 +\sqrt{\omega_1^2 + {\omega_0^2 - \omega_1^2\over 1 + e^{{T\over\d t}}}} \bigg)\bigg) + e^{T\over\d t}\omega_1^2\bigg(-1 + 4\d t\bigg(-\omega_1 +\sqrt{\omega_1^2 + {\omega_0^2 - \omega_1^2\over 1 + e^{{T\over\d t}}}} \bigg)\bigg) \bigg)\cr
&&\qquad\times{}_2F_1 (\d t (-\omega_0 +\omega_1), -\d t(\omega_0 + \omega_1 ) , 1 - 2\delta t\omega_0, -e^{{T\over\delta t}})\cr
&&\qquad + 4(1+e^{T\over\d t})\d t(\omega_0 + \omega_1)(\omega_0^2 + e^{T\over \d t}\omega_1^2){}_2F_1 (1-\d t (\omega_0 +\omega_1), \d t(-\omega_0 + \omega_1 ) , 1 - 2\delta t\omega_0, -e^{{T\over\delta t}})\bigg]\cr
&&\times\Bigg( \d t (\omega_0^2 + e^{T\over \d t}\omega_1^2)^2\bigg(\left(\omega _0-\omega _1\right)
   \, _2F_1\left(\text{$\delta $t} \left(\omega _0-\omega _1\right)+1,\text{$\delta $t} \left(\omega _0+\omega
   _1\right);2 \text{$\delta $t} \omega _0+1;-e^{T/\text{$\delta $t}}\right) \cr
   &&\qquad\qquad\qquad\qquad\quad _2F_1\left(\text{$\delta $t}
   \left(\omega _1-\omega _0\right),-\text{$\delta $t} \left(\omega _0+\omega _1\right);1-2 \text{$\delta $t} \omega
   _0;-e^{T/\text{$\delta $t}}\right)\cr
   &&\quad+\left(\omega _0+\omega _1\right) \, _2F_1\left(\text{$\delta $t} \left(\omega
   _0-\omega _1\right),\text{$\delta $t} \left(\omega _0+\omega _1\right);2 \text{$\delta $t} \omega
   _0+1;-e^{T/\text{$\delta $t}}\right) \cr
   && \qquad\qquad\qquad\qquad\quad_2F_1\left(\text{$\delta $t} \left(\omega _1-\omega
   _0\right),1-\text{$\delta $t} \left(\omega _0+\omega _1\right);1-2 \text{$\delta $t} \omega _0;-e^{T/\text{$\delta
   $t}}\right)\bigg)\Bigg)^{-1}
   \cr
   \cr
   \cr
   B' &=& {1\over 4}e^{T\omega_0-C(\d t,\omega_0,\omega_1)+F(T,\d t,\omega_0,\omega_1) }\bigg(\omega_1^2 + {\omega_0^2 - \omega_1^2\over 1 + e^{{T\over\d t}}}\bigg)^{3\over4}\cr
&&\bigg[\bigg(4\d t \omega_0^2\bigg(\omega_1 -\sqrt{\omega_1^2 + {\omega_0^2 - \omega_1^2\over 1 + e^{{T\over\d t}}}} \bigg) + 4e^{2{T\over\d t}}\d t \omega_1^2\bigg(\omega_1 -\sqrt{\omega_1^2 + {\omega_0^2 - \omega_1^2\over 1 + e^{{T\over\d t}}}} \bigg)\cr
&&\qquad + e^{T\over\d t}\omega_0^2\bigg(-1 + 4\d t\bigg(\omega_1 -\sqrt{\omega_1^2 + {\omega_0^2 - \omega_1^2\over 1 + e^{{T\over\d t}}}} \bigg)\bigg) + e^{T\over\d t}\omega_1^2\bigg(1 + 4\d t\bigg(\omega_1 -\sqrt{\omega_1^2 + {\omega_0^2 - \omega_1^2\over 1 + e^{{T\over\d t}}}} \bigg)\bigg) \bigg)\cr
&&\qquad\times{}_2F_1 (\d t (\omega_0 - \omega_1), \d t(\omega_0 + \omega_1 ) , 1 + 2\delta t\omega_0, -e^{{T\over\delta t}})\cr
&&\qquad + 4(1+e^{T\over\d t})\d t(\omega_0 - \omega_1)(\omega_0^2 + e^{T\over \d t}\omega_1^2){}_2F_1 (1+\d t (\omega_0 -\omega_1), \d t(\omega_0 + \omega_1 ) , 1 + 2\delta t\omega_0, -e^{{T\over\delta t}})\bigg]\cr
&&\times\Bigg( \d t (\omega_0^2 + e^{T\over \d t}\omega_1^2)^2\bigg(\left(\omega _0-\omega _1\right)
   \, _2F_1\left(\text{$\delta $t} \left(\omega _0-\omega _1\right)+1,\text{$\delta $t} \left(\omega _0+\omega
   _1\right);2 \text{$\delta $t} \omega _0+1;-e^{T/\text{$\delta $t}}\right) \cr
   &&\qquad\qquad\qquad\qquad\quad _2F_1\left(\text{$\delta $t}
   \left(\omega _1-\omega _0\right),-\text{$\delta $t} \left(\omega _0+\omega _1\right);1-2 \text{$\delta $t} \omega
   _0;-e^{T/\text{$\delta $t}}\right)\cr
   &&\quad+\left(\omega _0+\omega _1\right) \, _2F_1\left(\text{$\delta $t} \left(\omega
   _0-\omega _1\right),\text{$\delta $t} \left(\omega _0+\omega _1\right);2 \text{$\delta $t} \omega
   _0+1;-e^{T/\text{$\delta $t}}\right) \cr
   && \qquad\qquad\qquad\qquad\quad_2F_1\left(\text{$\delta $t} \left(\omega _1-\omega
   _0\right),1-\text{$\delta $t} \left(\omega _0+\omega _1\right);1-2 \text{$\delta $t} \omega _0;-e^{T/\text{$\delta
   $t}}\right)\bigg)\Bigg)^{-1}
   \label{u coeff}
\eea

\subsection{Solution for $v$}
Similarly for $v$ we write the following adiabatic solution
\bea
v_{ad}(\t) &=&{1\over\sqrt{\omega(\t)}} e^{C(\d t,\omega_0,\omega_1)-\int^{\t} \omega(\t')d\t'}\cr
&=&  {1\over\sqrt{\omega(\t)}}e^{C(\d t,\omega_0,\omega_1)-\int^{\t} \big(\omega_0^2 + {\omega_1^2-\omega_0^2\over 1 + e^{-{\t'\over\d t}}}\big)^{1\over2}d\t'}\cr
&=&  {1\over\sqrt{\omega(\t)}}e^{C(\d t,\omega_0,\omega_1)-F(\t,\d t,\omega_0,\omega_1)}
\label{adia v}
\eea
Similarly, as $\t \to - \infty$
\bea
v_{ad}(\t)\to {1\over\sqrt{\omega_0}}e^{-\omega_0\t}
\eea
\bea
C_v={1\over C_u}=e^{C(\d t,\omega_0,\omega_1)}
\eea
The derivative of $v_{ad}(\t)$ is given by.
\bea
\partial_{\t} v_{ad}(\t) &=& \bigg(-f(\t) - {\partial_{\t}f(\t)\over f(\t)} \bigg) v_{ad} (\t)
\label{der adia v}
\eea
Solving for the solution
\bea
v(\t) = A u'(\t) + B v'(\t)
\label{v}
\eea
we impose the boundary conditions
\bea
v(T) &=& v_{ad}(T)\cr
\partial_{\t} v(T)&=& \partial_{\t}v_{ad}(T)
\eea
for $T$ sufficiently less than $0$. This yields an $A$ and $B$ of
\bea
A&=&{1\over4}e^{-T\omega_0+C(\d t,\omega_0,\omega_1)-F(T,\d t,\omega_0,\omega_1) }\bigg(\omega_1^2 + {\omega_0^2 - \omega_1^2\over 1 + e^{{T\over\d t}}}\bigg)^{3\over4}\cr
&&\bigg[-\bigg(4\d t \omega_0^2\bigg(\omega_1 +\sqrt{\omega_1^2 + {\omega_0^2 - \omega_1^2\over 1 + e^{{T\over\d t}}}} \bigg) + 4e^{2{T\over\d t}}\d t \omega_1^2\bigg(\omega_1 +\sqrt{\omega_1^2 + {\omega_0^2 - \omega_1^2\over 1 + e^{{T\over\d t}}}} \bigg)\cr
&&\qquad + e^{T\over\d t}\omega_0^2\bigg(-1 + 4\d t\bigg(\omega_1 +\sqrt{\omega_1^2 + {\omega_0^2 - \omega_1^2\over 1 + e^{{T\over\d t}}}} \bigg)\bigg) + e^{T\over\d t}\omega_1^2\bigg(1 + 4\d t\bigg(\omega_1 +\sqrt{\omega_1^2 + {\omega_0^2 - \omega_1^2\over 1 + e^{{T\over\d t}}}} \bigg)\bigg) \bigg)\cr
&&\qquad\times{}_2F_1 (-\d t (\omega_0 +\omega_1), \d t(-\omega_0 + \omega_1 ) , 1 - 2\delta t\omega_0, -e^{{T\over\delta t}})\cr
&&\qquad + 4(1+e^{T\over\d t})\d t(\omega_0 + \omega_1)(\omega_0^2 + e^{T\over \d t}\omega_1^2){}_2F_1 (1-\d t (\omega_0 +\omega_1), \d t(-\omega_0 + \omega_1 ) , 1 - 2\delta t\omega_0, -e^{{T\over\delta t}})\bigg]\cr
&&\times\Bigg( \d t (\omega_0^2 + e^{T\over \d t}\omega_1^2)^2\bigg(\left(\omega _0-\omega _1\right)
   \, _2F_1\left(\text{$\delta $t} \left(\omega _0-\omega _1\right)+1,\text{$\delta $t} \left(\omega _0+\omega
   _1\right);2 \text{$\delta $t} \omega _0+1;-e^{T/\text{$\delta $t}}\right) \cr
   &&\qquad\qquad\qquad\qquad\quad _2F_1\left(\text{$\delta $t}
   \left(\omega _1-\omega _0\right),-\text{$\delta $t} \left(\omega _0+\omega _1\right);1-2 \text{$\delta $t} \omega
   _0;-e^{T/\text{$\delta $t}}\right)\cr
   &&\quad+\left(\omega _0+\omega _1\right) \, _2F_1\left(\text{$\delta $t} \left(\omega
   _0-\omega _1\right),\text{$\delta $t} \left(\omega _0+\omega _1\right);2 \text{$\delta $t} \omega
   _0+1;-e^{T/\text{$\delta $t}}\right) \cr
   && \qquad\qquad\qquad\qquad\quad_2F_1\left(\text{$\delta $t} \left(\omega _1-\omega
   _0\right),1-\text{$\delta $t} \left(\omega _0+\omega _1\right);1-2 \text{$\delta $t} \omega _0;-e^{T/\text{$\delta
   $t}}\right)\bigg)\Bigg)^{-1}
   \cr
   \cr
   \cr
   B&=&{1\over4}e^{T\omega_0+C(\d t,\omega_0,\omega_1)-F(T,\d t,\omega_0,\omega_1) }\bigg(\omega_1^2 + {\omega_0^2 - \omega_1^2\over 1 + e^{{T\over\d t}}}\bigg)^{3\over4}\cr
&&\bigg[\bigg(4\d t \omega_0^2\bigg(\omega_1 +\sqrt{\omega_1^2 + {\omega_0^2 - \omega_1^2\over 1 + e^{{T\over\d t}}}} \bigg) + 4e^{2{T\over\d t}}\d t \omega_1^2\bigg(\omega_1 +\sqrt{\omega_1^2 + {\omega_0^2 - \omega_1^2\over 1 + e^{{T\over\d t}}}} \bigg)\cr
&&\qquad + e^{T\over\d t}\omega_0^2\bigg(-1 + 4\d t\bigg(\omega_1 +\sqrt{\omega_1^2 + {\omega_0^2 - \omega_1^2\over 1 + e^{{T\over\d t}}}} \bigg)\bigg) + e^{T\over\d t}\omega_1^2\bigg(1 + 4\d t\bigg(\omega_1 +\sqrt{\omega_1^2 + {\omega_0^2 - \omega_1^2\over 1 + e^{{T\over\d t}}}} \bigg)\bigg) \bigg)\cr
&&\qquad\times{}_2F_1 (\d t (\omega_0 -\omega_1), \d t(\omega_0 + \omega_1 ) , 1 + 2\delta t\omega_0, -e^{{T\over\delta t}})\cr
&&\qquad + 4(1+e^{T\over\d t})\d t(\omega_0 - \omega_1)(\omega_0^2 + e^{T\over \d t}\omega_1^2){}_2F_1 (1+\d t (\omega_0 -\omega_1), \d t(\omega_0 + \omega_1 ) , 1 + 2\delta t\omega_0, -e^{{T\over\delta t}})\bigg]\cr
&&\times\Bigg( \d t (\omega_0^2 + e^{T\over \d t}\omega_1^2)^2\bigg(\left(\omega _0-\omega _1\right)
   \, _2F_1\left(\text{$\delta $t} \left(\omega _0-\omega _1\right)+1,\text{$\delta $t} \left(\omega _0+\omega
   _1\right);2 \text{$\delta $t} \omega _0+1;-e^{T/\text{$\delta $t}}\right) \cr
   &&\qquad\qquad\qquad\qquad\quad _2F_1\left(\text{$\delta $t}
   \left(\omega _1-\omega _0\right),-\text{$\delta $t} \left(\omega _0+\omega _1\right);1-2 \text{$\delta $t} \omega
   _0;-e^{T/\text{$\delta $t}}\right)\cr
   &&\quad+\left(\omega _0+\omega _1\right) \, _2F_1\left(\text{$\delta $t} \left(\omega
   _0-\omega _1\right),\text{$\delta $t} \left(\omega _0+\omega _1\right);2 \text{$\delta $t} \omega
   _0+1;-e^{T/\text{$\delta $t}}\right) \cr
   && \qquad\qquad\qquad\qquad\quad_2F_1\left(\text{$\delta $t} \left(\omega _1-\omega
   _0\right),1-\text{$\delta $t} \left(\omega _0+\omega _1\right);1-2 \text{$\delta $t} \omega _0;-e^{T/\text{$\delta
   $t}}\right)\bigg)\Bigg)^{-1}
    \label{v coeff}
\eea


\subsection{Solutions for $\rho^2$}
Let us check that the set of adiabatic initial conditions chosen for $u(\t)$ and $v(\t)$ in (\ref{adia u}), (\ref{der adia u}), (\ref{adia v}), and (\ref{der adia v}) and thus $\rho^2(\t)$ in (\ref{rho sq 2}) are consistent with choosing adiabatic initial conditions for $\rho^2(\t)$ without reference to $u(\t)$ and $v(\t)$. Since the equation for $\rho^2(\t)$ is (\ref{rho sq 2}), when $\t\to T$ we have the first initial condition
\bea
\rho^2(\t)\to u_{ad}(T)v_{ad}(T)={1\over f(T)}=\rho^2_{ad}(T)
\eea
where 
\bea
\rho^2_{ad}(\t) \equiv{1\over f(\t)}
\eea
The second initial condition uses $\partial_{\t}\rho^2(\t)$ where
\bea
\partial_{\t} \rho^2(\t) &=&v(\t) \partial_{\t}  u(\t)    +  u(\t)  \partial_{\t} v(\t)
\eea
Taking $\t \to T$ yields
\bea
\partial_{\t} \rho^2(\t)&\to& v_{ad}(T) \partial_{\t}  u(T)    +  u_{ad}(T)  \partial_{\t} v (T)\cr
&=&v_{ad}(T) \partial_{\t}  u_{ad}(T)    +  u_{ad}(T)  \partial_{\t} v_{ad} (T)\cr
&=&\bigg(f(T) - {\partial_{\t}f(T)\over2f(T)} \bigg)u_{ad}(T) v_{ad}(T)  + \bigg(-f(T) - {\partial_{\t}f(T)\over2f(T)} \bigg)u_{ad} (T)v_{ad}(T)\cr
&=&-{\partial_{\t}f(T)\over f(T)^2} \cr
&=&\partial_{\t} \rho_{ad}^2(T)
\eea
where 
\bea
\partial_{\t} \rho_{ad}^2(\t) = -{\partial_{\t}f(\t)\over f(\t)^2}
\eea
Therefore, we see that the set of initial conditions for $u(\t)$ and $v(\t)$ respectively are consistent with choosing $\rho^2(T)=\rho_{ad}^2(T)={1\over f(T)}$ and $\partial_{\t} \rho^2(T)=\partial_{\t} \rho_{ad}^2(T)=-{\partial_{\t}f(T)\over f(T)^2}$  without referring to the functions $u(\t),v(\t)$. Therefore the solution for $\rho^2(\t)$ is 
\bea
\rho^2(\t) = u(\t)v(\t)
\eea 
with $u(\t)$ and $v(\t)$ given in (\ref{u}), (\ref{v}), (\ref{u coeff}), (\ref{v coeff}) and (\ref{coeff}).


\section{Smooth Pulse and Dip protocols $\omega_0\to\omega_1\to\omega_0$}\label{smooth ccp}
Performing similar computations as for the smooth step protocols in the previous section we solve for $\rho^2(\t)$ for a smooth pulse for $\omega_0\to\omega_1\to\omega_0$. The $f(\t)^2$ is given by 

\bea
f(\t)^2  &=& \omega_1^2+ (\omega_0^2 - \omega_1^2)\tanh^2 {\t\over\d t}
\label{smooth CCP profile}
\eea
We again apply adiabatic initial conditions for $u(\t)$ and $v(\t)$ at some early time $T$ sufficiently smaller than $0$. We obtain
\bea
\rho^2(\t) = u(\t)v(\t)
\eea
The solutions for $u,v$ are given
\bea
u(\t)&=& A' u'(\t) + B' v'(\t)\cr
v(\t) &=& A u'(\t) + B v'(\t)
\eea
with
\bea
u'(\t) &=& P^{\mu}_{\nu}(\tanh{\t\over\d t})\cr
v'(\t) &=& Q^{\mu}_{\nu}(\tanh{\t\over\d t})
\eea
and
\bea
\mu &=& \omega_0\d t\cr
\nu&=& {1\over2}(-1+\sqrt{1+4\d t^2(\omega_0^2-\omega_1^2)})
\label{mu nu 2}
\eea
$u'(\t)$ and $v'(\t)$ are solutions of
\bea
{d^2u'\over d\tau^2 } -f(\t)^2 u' &=&0\cr
{d^2v'\over d\tau^2 } -f(\t)^2 v' &=&0
\eea
with $f(\t)$ given in (\ref{smooth CCP profile}). $P^{\mu}_{\nu}(z)$ and $Q^{\mu}_{\nu}(z)$ are associated Legendre Polynomials of the first and second kind respectively. The coefficients $A',B',A,B$ are given by the expressions
\bea
A'&=&-e^{\tilde F(T,\d t,\omega_0,\omega_1)}\cr
&&\bigg[- \bigg( 1 - 2 \d t \omega_0 + \sqrt{1 + 4 \d t^2(\omega_0^2 -\omega_1^2 )}\bigg)\bigg(-\omega_0^2 + (\omega_0^2 - \omega_1^2 ) \text{sech}^2 \bigg({T\over \d t}\bigg)\bigg)Q_{1+\nu}^{\mu}(\text{tanh}{T\over\d t})\cr
&&+\bigg(-\omega_0^2\bigg(1 + \sqrt{1 + 4\d t^2(\omega_0^2 - \omega_1^2)}\bigg)\text{tanh}{T\over\d t} + \text{sech}^2 {T\over \d t} \bigg( -\d t (\omega_0^2 - 2 \omega_1^2)\sqrt{\omega_0^2+ (\omega_1^2-\omega_0^2 )\text{sech}^2{T\over\d t}}\cr
&&\qquad\quad+\d t\omega_0^2\text{cosh} {2T\over\d t}\sqrt{\omega_0^2 + (\omega_1^2 - \omega_0^2)\text{sech}^2{T\over \d t}}+( \omega_0^2-\omega_1^2)\sqrt{1+4\d t^2(\omega_0^2-\omega_1^2)}\text{tanh}{T\over\d t} \bigg)\bigg)Q_{\nu}^{\mu}(\text{tanh}{T\over\d t}) \bigg]\cr
&&\times \bigg[  \bigg(1- 2 \d t\omega_0 + \sqrt{1 + 4 \d t^2(\omega_0^2 - \omega_1^2)}\bigg)\bigg(\omega_0^2 + (- \omega_0^2 + \omega_1^2 )\text{sech}^2{T\over\d t}\bigg)^{5\over4}\cr
&&\qquad\qquad\bigg( P_{1+\nu}^{\mu}(\tanh {T\over\d t})Q_{\nu}^{\mu}(\text{tanh} {T\over\d t}) - P_{\nu}^{\mu}(\text{tanh} {T\over\d t})Q_{1+\nu}^{\mu}(\text{tanh} {T\over\d t})\bigg)\bigg]^{-1}
\cr
\cr
\cr
B'&=&-e^{\tilde F(T,\d t,\omega_0,\omega_1)}\text{sech}^2{T\over\d t}\cr
&&\bigg[2 \bigg( 1 - 2 \d t \omega_0 + \sqrt{1 + 4 \d t^2(\omega_0^2 -\omega_1^2 )}\bigg)\bigg(-\omega_0^2 +2 \omega_1^2 + \omega_0^2 \cosh \bigg({2T\over \d t}\bigg)\bigg)P_{1+\nu}^{\mu}(\tanh{T\over\d t})\cr
&&+\bigg(-\omega_0^2\text{sinh}{2T\over\d t} - (\omega_0^2-2\omega_1^2)\bigg(-\sqrt{1 + 4\d t^2(\omega_0^2 - \omega_1^2)}\tanh{T\over\d t} + 2\d t \sqrt{\omega_0^2+ (\omega_1^2-\omega_0^2 )\text{sech}^2{T\over\d t}}\bigg)\cr
&&\qquad\quad+\omega_0^2\cosh {2T\over\d t}\bigg(- \sqrt{1 + 4\d t^2(\omega_0^2 - \omega_1^2)}\tanh{T\over\d t} + 2\d t \sqrt{\omega_0^2+ (\omega_1^2-\omega_0^2 )\text{sech}^2{T\over\d t}}\bigg) \bigg)2P_{\nu}^{\mu}(\tanh{T\over\d t}) \bigg]\cr
&&\times \bigg[  4\bigg(1- 2 \d t\omega_0 + \sqrt{1 + 4 \d t^2(\omega_0^2 - \omega_1^2)}\bigg)\cr
&&\qquad\qquad\bigg( P_{1+\nu}^{\mu}(\tanh {T\over\d t})Q_{\nu}^{\mu}(\tanh {T\over\d t}) - P_{\nu}^{\mu}(\tanh {T\over\d t})Q_{1+\nu}^{\mu}(\tanh {T\over\d t})\bigg)\bigg]^{-1}
\eea

\bea
A&=&e^{-\tilde F(T,\d t,\omega_0,\omega_1)}\text{sech}^2 {T\over \d t}\cr
&&\bigg[-2 \bigg( 1 - 2 \d t \omega_0 + \sqrt{1 + 4 \d t^2(\omega_0^2 -\omega_1^2 )}\bigg)\bigg(-\omega_0^2 + 2\omega_1^2 +\omega_0^2 \text{cosh} \bigg({2T\over \d t}\bigg)\bigg)Q_{1+\nu}^{\mu}(\text{tanh}{T\over\d t})\cr
&&+\bigg(-2\d t \big(\omega_0^2 - 4 \omega_1^2)\text{cosh}{T\over \d t}\sqrt{\omega_0^2 + (\omega_1^2 - \omega_0^2)\text{sech}^2{T\over \d t}}+ 2 \d t\omega_0^2\text{cosh} {3T\over\d t}\sqrt{\omega_0^2 + (\omega_1^2 - \omega_0^2)\text{sech}^2 {T\over \d t}}\cr
&&\qquad + 2\text{sinh}{T\over\d t}\bigg( - (\omega_0^2 - 2\omega_1^2)\sqrt{1+ 4 \d t^2 (\omega_0^2 - \omega_1^2)} + \omega_0^2\bigg(1 + \bigg(1+\sqrt{1+ 4 \d t^2 (\omega_0^2 - \omega_1^2)}\bigg)\text{cosh}{2T\over\d t}\bigg)\bigg)\bigg)\cr
&&\qquad \times\text{sech}{T\over\d t} Q_{\nu}^{\mu}(\text{tanh}{T\over\d t}) \bigg]\cr
&&\times \bigg[ 4 \bigg(1- 2 \d t\omega_0 + \sqrt{1 + 4 \d t^2(\omega_0^2 - \omega_1^2)}\bigg)\bigg(\omega_0^2 + (\omega_1^2 - \omega_0^2)\text{sech}^2{T\over\d t}\bigg)^{5\over4}\cr
&&\qquad\qquad\bigg( P_{1+\nu}^{\mu}(\text{tanh} {T\over\d t})Q_{\nu}^{\mu}(\text{tanh} {T\over\d t}) - P_{\nu}^{\mu}(\text{tanh} {T\over\d t})Q_{1+\nu}^{\mu}(\text{tanh} {T\over\d t})\bigg)\bigg]^{-1}
   \cr
   \cr
   B&=& -e^{-\tilde F(T,\d t,\omega_0,\omega_1)}\text{sech}^2 {T\over \d t}\cr
&&\bigg[-2 \bigg( 1 - 2 \d t \omega_0 + \sqrt{1 + 4 \d t^2(\omega_0^2 -\omega_1^2 )}\bigg)\bigg(-\omega_0^2 + 2\omega_1^2 +\omega_0^2 \text{cosh} \bigg({2T\over \d t}\bigg)\bigg)P_{1+\nu}^{\mu}(\text{tanh}{T\over\d t})\cr
&&+\bigg(-2\d t \big(\omega_0^2 - 4 \omega_1^2)\text{cosh}{T\over \d t}\sqrt{\omega_0^2 + (\omega_1^2 - \omega_0^2)\text{sech}^2{T\over \d t}}+ 2 \d t\omega_0^2\text{cosh} {3T\over\d t}\sqrt{\omega_0^2 + (\omega_1^2 - \omega_0^2)\text{sech}^2 {T\over \d t}}\cr
&&\qquad + 2\text{sinh}{T\over\d t}\bigg( - (\omega_0^2 - 2\omega_1^2)\sqrt{1+ 4 \d t^2 (\omega_0^2 - \omega_1^2)} + \omega_0^2\bigg(1 + \bigg(1+\sqrt{1+ 4 \d t^2 (\omega_0^2 - \omega_1^2)}\bigg)\text{cosh}{2T\over\d t}\bigg)\bigg)\bigg)\cr
&&\qquad \times\text{sech}{T\over\d t} P_{\nu}^{\mu}(\text{tanh}{T\over\d t}) \bigg]\cr
&&\times \bigg[ 4 \bigg(1- 2 \d t\omega_0 + \sqrt{1 + 4 \d t^2(\omega_0^2 - \omega_1^2)}\bigg)\bigg(\omega_0^2 + (\omega_1^2 - \omega_0^2)\text{sech}^2{T\over\d t}\bigg)^{5\over4}\cr
&&\qquad\qquad\bigg( P_{1+\nu}^{\mu}(\text{tanh} {T\over\d t})Q_{\nu}^{\mu}(\text{tanh} {T\over\d t}) - P_{\nu}^{\mu}(\text{tanh} {T\over\d t})Q_{1+\nu}^{\mu}(\text{tanh} {T\over\d t})\bigg)\bigg]^{-1}
   \eea
where
\bea
\tilde F(T,\d t,\omega_0,\omega_1)=\d t \bigg(\omega_0 \text{sinh}^{-1}({\omega_0\over\omega_1}\text{sinh}({T\over\d t})) + \sqrt{\omega_1^2 - \omega_0^2}\text{tan}^{-1} \bigg({\sqrt{-2\omega_0^2 + 2\omega_1^2}\text{sinh}{T\over \d t}\over\sqrt{-\omega_0^2 +2\omega_1^2 +\omega_0^2\text{cosh}({2T\over\d t})}}\bigg)  \bigg)\nn
\eea

\section{Phase space density for a potential with a cutoff}
\label{apppsd}
In this part we explain in detail the fermi surface profiles for a potential with a cutoff at $x = \pm l/2$. Without loss of generality, we consider coordinate $(x,p)$ on the left and up side of phase space.

After abrupt quench, the equi-energy trajectories become 
\begin{equation}
\frac{1}{2}p^2-\frac{1}{2}\omega_1^2 x^2=E 
\end{equation}
which can be described by hyperbolic functions
\begin{equation}
\left\{ 
\begin{array}{*{20}{c}}
x= \frac{\sqrt{2E}}{\omega_1} \sinh (\omega_1 \tau + \phi), \\
p=\sqrt{2E} \cosh (\omega_1 \tau +\phi), \\
\end{array}
\right. (E>0)
\&
\left\{ 
\begin{array}{*{20}{c}}
x=- \frac{\sqrt{-2E}}{\omega_1} \cosh (\omega_1 \tau + \phi), \\
p=- \sqrt{-2E} \sinh (\omega_1 \tau +\phi), \\
\end{array}
\right. (E<0)
\end{equation} 
Here the energy $E$ and phase $\phi$ are two independent parameters determined by the continuity of phase space trajectories, i.e. the continuity of $(x,p)$ during the abrupt quench.

Now, because the potential has cutoffs at $x= \pm l/2$, all the classical particles should be bounced back. This implies that the $(x=\pm l/2,p)$ and $(x=\pm l/2,-p)$ are identical in phase space. In terms of $E$ and $\phi$, the reflection of particles at the boundary only change the phase $\phi$, and the coordinates of the particle in phase space after reflection can be described by
\begin{equation}
\left\{ 
\begin{array}{*{20}{c}}
x= -\frac{\sqrt{2E}}{\omega_1} \sinh (\omega_1 (\tau-\tau_l) -\phi_l), \\
p=- \sqrt{2E} \cosh (\omega_1 (\tau-\tau_l) -\phi_l) \\
\end{array}
\right. (E>0)
\&
\left\{ 
\begin{array}{*{20}{c}}
x=- \frac{\sqrt{-2E}}{\omega_1} \cosh (\omega_1 (\tau- \tau_l) - \phi_l), \\
p=- \sqrt{-2E} \sinh (\omega_1 (\tau-\tau_l) -\phi_l), \\
\end{array}
\right. (E<0)
\label{xp_r}
\end{equation} 
where $\tau_l$ is the time the particle needs to reach the boundary to be reflected, and $\phi_l$ is the phase of the particle at the boundary before reflection (after reflection the phase becomes $-\phi_l$). Thus they satisfy the relation
\begin{equation}
\phi_l =\omega_1 \tau_l +\phi
\end{equation}
Therefore, we can simplify (\ref{xp_r}) and find that the reflection is equivalent to the transformation on $E$ and $\phi$:
\begin{equation}
E \to E,~~~ \& ~~~ \phi \to \phi-2\phi_l
\end{equation} 
Now we can rewrite the coordinates of the particle that bounced back as
\begin{equation}
\begin{split}
\left\{ 
\begin{array}{*{20}{c}}
x_b=-x\cosh 2\phi_l + \frac{1}{\omega_1} p \sinh 2\phi_l \\
p_b=-p\cosh 2\phi_l + \omega_1 x \sinh 2\phi_l \\
\end{array}
\right. (E>0) \\
\&
\left\{ 
\begin{array}{*{20}{c}}
x_b= x \cosh 2\phi_l - \frac{1}{\omega_1}p \sinh 2\phi_l \\
p_b= p \cosh 2\phi_l - \omega_1 x \sinh 2\phi_l , \\
\end{array}
\right. (E<0)
\end{split}
\end{equation} 
which implies a boost 
\begin{equation}
K = \left( 
\begin{array}{*{20}{c}}
\cosh 2 \phi_l & -\sinh 2\phi_l \\
-\sinh 2\phi_l & \cosh 2\phi_l \\
\end{array}
\right)
\end{equation}
from $(x, p/\omega_1)$ to $(x_b, p_b/\omega_1)$. For $E>0$, the boost is $-K$; for $E<0$, it is $K$. The boost can be applied to the coordinates of the particles after the $n$-th reflection to obtain the coordinates after the $(n+1)$-th reflection. The inverse boost is
\begin{equation}
K^{-1}=  \left( 
\begin{array}{*{20}{c}}
\cosh 2 \phi_l &  \sinh 2\phi_l \\
 \sinh 2\phi_l & \cosh 2\phi_l \\
\end{array}
\right)
\end{equation}
and therefore,
\begin{equation}
\begin{split}
\left\{ 
\begin{array}{*{20}{c}}
x=-x_b\cosh 2\phi_l - \frac{1}{\omega_1} p_b \sinh 2\phi_l \\
p=-p_b\cosh 2\phi_l - \omega_1 x_b \sinh 2\phi_l \\
\end{array}
\right. (E>0) \\
\&
\left\{ 
\begin{array}{*{20}{c}}
x= x_b \cosh 2\phi_l + \frac{1}{\omega_1}p_b \sinh 2\phi_l \\
p= p_b \cosh 2\phi_l + \omega_1 x_b \sinh 2\phi_l , \\
\end{array}
\right. (E<0)
\end{split}
\label{xp_ri}
\end{equation} 
We can figure out the matrix element from $x=\pm l/2$
\begin{equation}
\begin{split}
\left(\frac{l}{2}\right)^2=\frac{2E}{\omega_1^2} \sinh^2 \phi_l \Longrightarrow \cosh 2\phi_l =1+2\sinh^2 \phi_l = 1+ \frac{\omega_1^2 l^2}{4E},~~~ (E>0) \\
\left(\frac{l}{2}\right)^2=\frac{-2E}{\omega_1^2} \cosh^2 \phi_l \Longrightarrow \cosh 2\phi_l =2\cosh^2 \phi_l-1 = -1 -\frac{\omega_1^2 l^2}{4E}.~~~ (E<0)
\end{split}
\end{equation}

Notice that the Fermi surface for an unbounded potential satisfies (\ref{4-7}), we can plug (\ref{xp_ri}) in and obtain the relation satisfied by particle reflected back once:
\begin{equation}
\begin{split}
&\left(  x \cosh 2\phi_l + \frac{1}{\omega_1}p \sinh 2\phi_l \right)^2 \\
&~~~ -\left[\rho(\tau)^2 \left(  p \cosh 2\phi_l + \omega_1 x \sinh 2\phi_l \right) -\frac{1}{2}\left(  x \cosh 2\phi_l + \frac{1}{\omega_1}p \sinh 2\phi_l \right) \partial_{\tau} \rho^2 \right]^2 = \rho(\tau)^2
\end{split}
\label{fs_r}
\end{equation}
Here we have omitted the lower index "$b$" since (\ref{fs_r}) represents part of the Fermi surface at $\tau$. The complete Fermi surface after quench is a combination of (\ref{fs_r}) and (\ref{4-7}) and $x=\pm l/2$. 

Finally we consider two cases: $\omega_0 > \omega_1$ and $\omega_0<\omega_1$. For $\omega_0 > \omega_1$, more particles flow in from the upper side (where $E>0$) than flow out from the lower side (where $E>0$) when there is no cutoff. This implies that when there is a cutoff, less particles should flow in and therefore, the Fermi surface due to reflection should give a third bound for the Fermi surface. As a result, the phase space density should be
\begin{equation}
\begin{split}
u=&\Theta \left( \frac{x^2}{\rho^2} - \frac{1}{\rho^2}\left( p \rho^2 -\frac{1}{2} x \partial_{\tau} \rho^2 \right)^2 -1 \right) \\
&\Theta \left( \frac{1}{\rho^2}\left(  x \cosh 2\phi_l + \frac{1}{\omega_1}p \sinh 2\phi_l \right)^2\right. \\
&~~~~~~\left. - \frac{1}{\rho^2}\left( \left[p \cosh 2\phi_l + \omega_1 x \sinh 2\phi_l \right] \rho^2 -\frac{1}{2} \left[ x \cosh 2\phi_l + \frac{1}{\omega_1}p \sinh 2\phi_l \right] \partial_{\tau} \rho^2 \right)^2 -1 \right) 
\end{split}
\end{equation}
where $\cosh 2\phi_l = 1+ \frac{\omega_1^2 l^2}{4E}$, $E>0$. The phase space density is shown in figure \ref{psds}.

\begin{figure}[H]
\centering
\subfloat[$\omega_0=2\omega_1=1$]{
\includegraphics[width=0.4\textwidth]{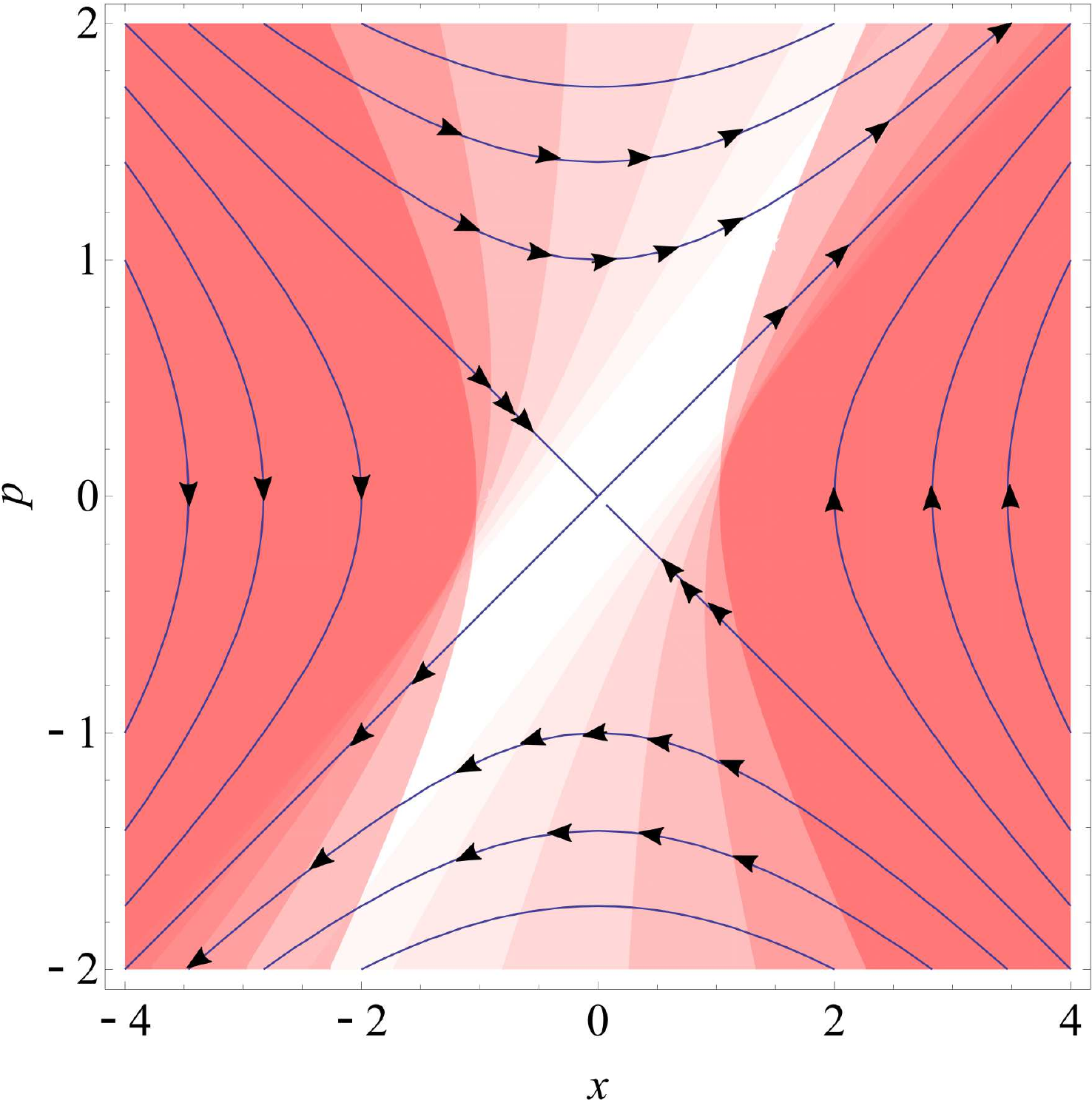}
\includegraphics[width=0.4\textwidth]{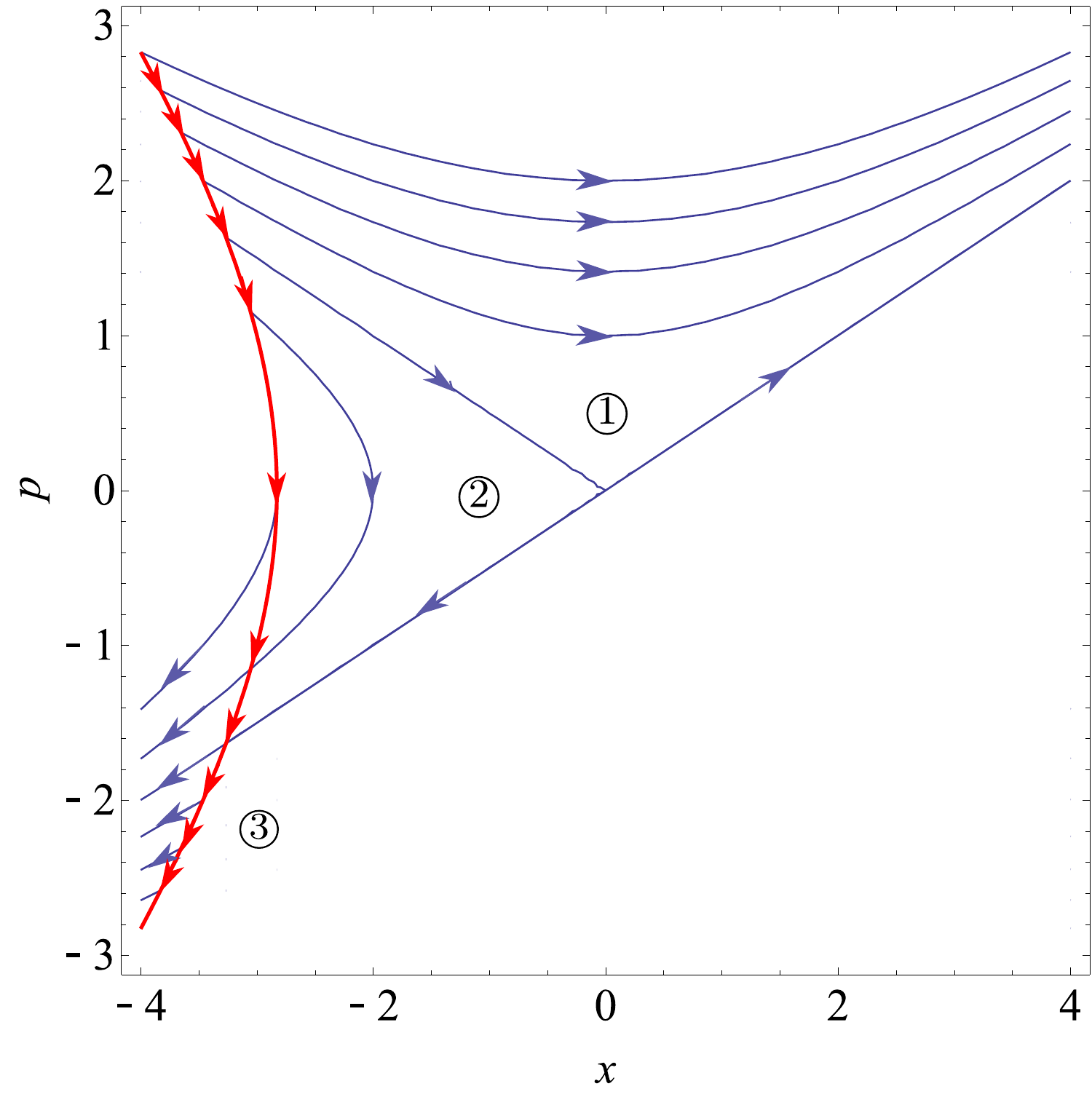}}\\
\subfloat[$\omega_0= \omega_1/2=1$]{
\includegraphics[width=0.4\textwidth]{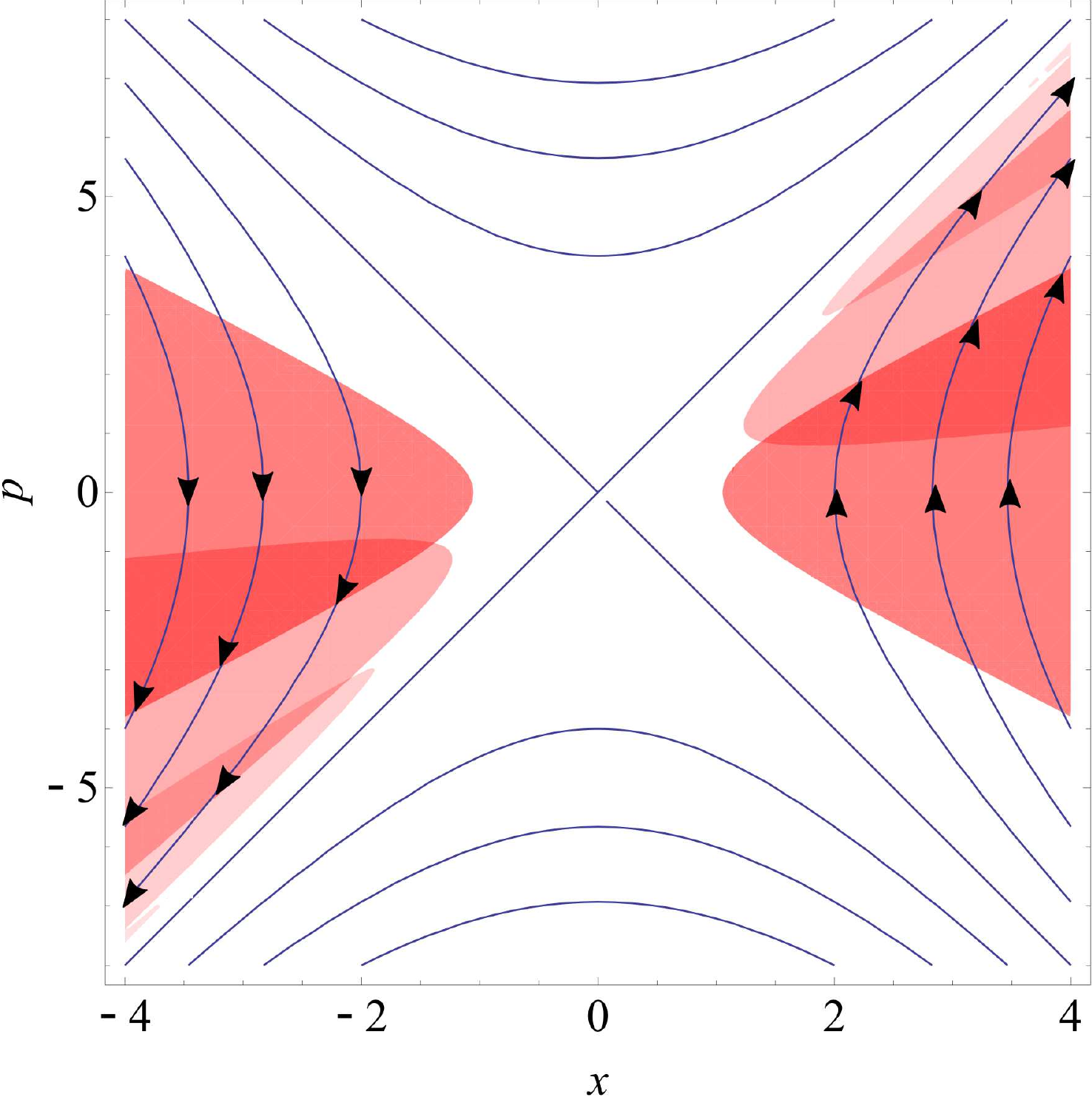}
\includegraphics[width=0.4\textwidth]{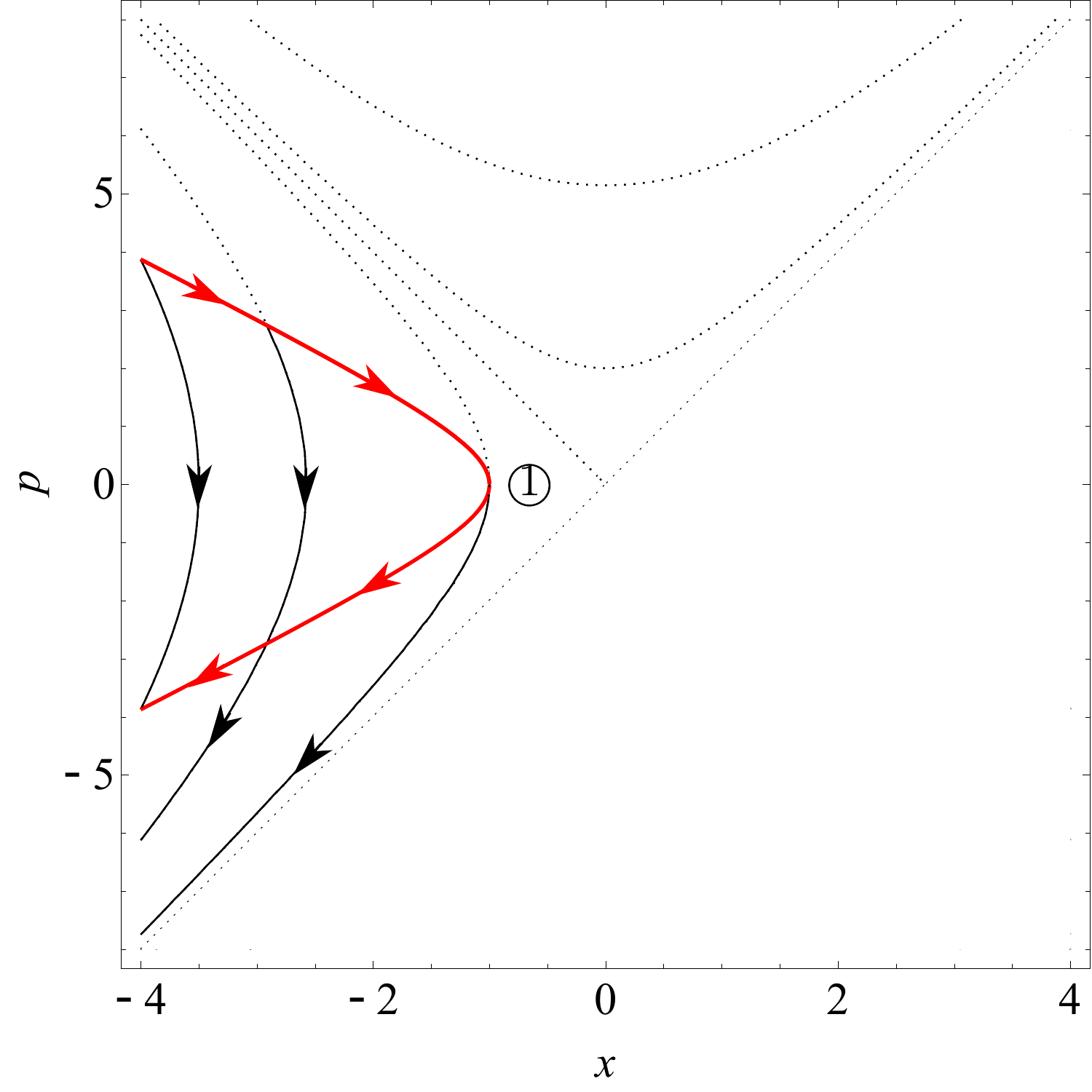}}
\caption{The formation of Fermi surface (\ref{4-7}). Left: the Fermi surfaces are plotted more transparently as time evolves. Solid lines are equi-energy surfaces i.e. the trajectories of classical particles after abrupt quench. The directions of particle motion are labeled by arrows. Right: the original Fermi surface when $\tau \le 0$ is plotted in solid red with arrows pointing in the directions of classical particles. Blue or black solid lines are the trajectories of classical particles after abrupt quench. The orientations of motion are labeled by arrows. Dashed black lines are equi-energy surfaces.}
\end{figure}

For $\omega_0 < \omega_1$, no particles flow in from the upper side (where $E<0$) while many particles flow out from the lower side (where $E<0$) when there is no cutoff. Therefore, when there is a cutoff, particles reflected should flow in from the upper side and occupy the empty space. As a result, the phase space density should be
\begin{equation}
\begin{split}
u=&\Theta \left( \frac{x^2}{\rho^2} - \frac{1}{\rho^2}\left( p \rho^2 -\frac{1}{2} x \partial_{\tau} \rho^2 \right)^2 -1 \right) \\
&+\Theta \left( \frac{1}{\rho^2}\left(  x \cosh 2\phi_l + \frac{1}{\omega_1}p \sinh 2\phi_l \right)^2\right. \\
&~~~~~~\left. - \frac{1}{\rho^2}\left( \left[p \cosh 2\phi_l + \omega_1 x \sinh 2\phi_l \right] \rho^2 -\frac{1}{2} \left[ x \cosh 2\phi_l + \frac{1}{\omega_1}p \sinh 2\phi_l \right] \partial_{\tau} \rho^2 \right)^2 -1 \right) \\
-
&\Theta \left( \frac{x^2}{\rho^2} - \frac{1}{\rho^2}\left( p \rho^2 -\frac{1}{2} x \partial_{\tau} \rho^2 \right)^2 -1 \right) \\
&\times \Theta \left( \frac{1}{\rho^2}\left(  x \cosh 2\phi_l + \frac{1}{\omega_1}p \sinh 2\phi_l \right)^2\right. \\
&~~~~~~\left. - \frac{1}{\rho^2}\left( \left[p \cosh 2\phi_l + \omega_1 x \sinh 2\phi_l \right] \rho^2 -\frac{1}{2} \left[ x \cosh 2\phi_l + \frac{1}{\omega_1}p \sinh 2\phi_l \right] \partial_{\tau} \rho^2 \right)^2 -1 \right) 
\end{split}
\end{equation}
where $\cosh 2\phi_l = -1 -\frac{\omega_1^2 l^2}{4E}$, $E<0$. The phase space density is shown in figure \ref{psdl}.

\end{appendix}

\end{document}